\documentclass[aps,pre,superscriptaddress,showpacs, reprint,amsmath,  amssymb,nofootinbibi]{revtex4-1}

\usepackage{amsmath}
\usepackage{amsthm}
\usepackage{amsfonts}
\usepackage{amssymb}
\usepackage{graphicx}
\usepackage{comment}

\usepackage{mathalfa}
\usepackage{psfrag}
\usepackage{xcolor}

\usepackage[all]{xy}

\newcommand{\one}{($i$) }
\newcommand{\two}{($ii$) }
\newcommand{\three}{($iii$) }

\usepackage{mathtools}
\DeclarePairedDelimiter{\ceil}{\lceil}{\rceil}

\usepackage{array}
\usepackage{comment}
\usepackage{mathtools}

\usepackage{tikz}
\usetikzlibrary{arrows,decorations.pathmorphing,backgrounds,positioning,fit,matrix}
\usetikzlibrary{shapes.arrows}
\usetikzlibrary{shadows.blur}
\usetikzlibrary{decorations.pathreplacing,shapes.misc}

\usepackage{subcaption}

\begin{document}

\title{Information ratchets exploiting spatially structured information reservoirs}

\author{Richard E. Spinney}
\email{richard.spinney@sydney.edu.au}
\affiliation{Complex Systems Research Group and Centre for Complex Systems, The University of Sydney, NSW 2006, Australia}
\author{Mikhail Prokopenko}

\affiliation{Complex Systems Research Group and Centre for Complex Systems, The University of Sydney, NSW 2006, Australia}
\author{Dominique Chu}

\affiliation{School of Computing, University of Kent, CT2 7NF, Canterbury, UK}
\date{\today}

\begin{abstract}
Fully mechanized ``Maxwell's demons,'' sometimes also called {\em information ratchets} are an important conceptual link between computation, information theory and statistical physics. They  exploit low entropy information reservoirs  to extract work from a heat reservoir.  Previous models of such demons have either ignored the cost of delivering bits to the demon from the information reservoir or assumed  random access or ``infinite dimensional'' information reservoirs to avoid such an issue.  In this work we account for this cost when exploiting information reservoirs with physical structure and show that the dimensionality of the reservoir has a significant impact on the performance and phase diagram of the demon. We find that  for conventional 1-dimensional tapes  the  scope for work extraction is greatly reduced. An expression for the net-extracted work by demons exploring information reservoirs by means of biased random walks on $d$-dimensional, $\mathbb{Z}^d$, information reservoirs is presented. Furthermore, we  derive exact probabilities of recurrence in these systems, generalizing previously known results. We  find that the demon is characterized by  two critical dimensions. Firstly to extract work at zero bias the dimensionality of the information reservoir must be larger than  $d=2$ corresponding to the dimensions where a simple random walker is transient. Secondly, for integer dimensions $d>4$ the unbiased random walk optimizes work extraction corresponding to the dimensions where a simple random walker is strongly transient. 

\end{abstract}

\maketitle

\section{Introduction.}
Through frameworks such as stochastic thermodynamics \cite{seifert_stochastic_2008,seifert_stochastic_2012} it has become clear that  order  can be treated as a thermodynamic  resource \cite{deffner_information_2013,barato_unifying_2014}, thus relating information theory, computation and statistical mechanics  \cite{sagawa_second_2008,horowitz_nonequilibrium_2010,toyabe_experimental_2010,esposito_stochastic_2012,sagawa_nonequilibrium_2012,horowitz_thermodynamics_2014,horowitz_secondlawlike_2014,barato_unifying_2014,strasberg_thermodynamics_2013,sagawa_fluctuation_2012}. A useful theoretical tool to probe the connection between information and thermodynamics are    {\em information ratchets}. These are     mechanized Maxwell's demons \cite{chapman_how_2015,m._d._vidrighin_photonic_2016}, that    extract work from a single heat reservoir while randomizing  reservoirs of low entropy, so-called   {\em information reservoirs}. 
\par
 A pioneering model of an information ratchet  is due to    Mandal and Jarzynski (MJ) \cite{mandal_work_2012}. Their ratchet contains three internal states and is in contact with an external tape that contains $1$s and $0$s and slides past the demon at a constant speed. At any one time, the ratchet is in contact with,  or bound to, a symbol on the tape.  The demon also performs thermally driven internal state transitions, which  are coupled to flips of the symbols on the currently bound tape element.  Due to the motion of the tape, the demon will occasionally get into contact with a new tape element, which may take it  out of equilibrium and induce  a cyclic sequence of state transitions. This  can be exploited to extract  energy from the thermal environment and store it as work in a work reservoir while randomizing the tape.   Barato and Seifert \cite{barato_stochastic_2014,barato_unifying_2014} (BS) (alongside other extensions \cite{hoppenau_energetics_2014,barato_autonomous_2013,mandal_maxwells_2013,rana_multipurpose_2016}) later simplified  the MJ model  to a fully stochastic  2-state model  with two crucial differences:  They replaced the  1D tape of the MJ model  by an infinitely large, spatially unstructured 
 (effectively infinite dimensional) 
 information reservoir behaving as a  perfectly mixed symbol gas. A consequence of this assumption is that the demon can transition from any reservoir element to any other reservoir element in one step, thus removing the need to assume  constant motion of a tape.  They also simplified the demon   to a two state model, such that at any time the demon is thermally fluctuating between a high and a low energy state. These transitions are again coupled to flips of the symbol associated with the currently bound element of the information reservoir. As the BS demon gets in contact with a new element of the information reservoir, the system may be taken out of equilibrium and the ensuing relaxation process can be exploited to extract work from the system while randomizing the  outgoing reservoir elements.  
\par
In this contribution we show that the long-term work extraction rate of a demon  is crucially dependent on  the geometric properties of the  information reservoir. We find that  information reservoirs without spatial structure,  as with the BS model, are  optimal  for work extraction. In contrast,  low dimensional reservoirs, such as  the canonical case of a one-dimensional tape, support work extraction  only  for a small range of demon parameters, and even then the best possible net-gain of work is small compared to  what is achievable in the unstructured case. Specifically, by considering the resultant dynamics and thermodynamics of the implicit mechanism which enables exploration of the information reservoir, we will demonstrate how the performance of such a demon varies between information reservoirs represented as $d$-dimensional cubic lattices, ranging from 1D tapes to infinite dimensional structures which attain random access.  We find two critical dimensions that separate qualitatively different regimes for exploitation of information reservoirs.  As such our results provide powerful insight into the physical and geometric factors which will arise in systems which use a persistent physical substrate to encode information.
\par
The understanding of such machines is not just of conceptual value. Examples of information reservoirs can be readily found in biology.   Already in the 1980s Bennet \cite{bennett_thermodynamics_1982} suggested the possibility of biological information ratchets that extract work from correlations between polymers such as DNA, mRNA or amino acid sequences; a concrete design was recently proposed by McGrath and coworkers \cite{mcgrath_biochemical_2017}. Yet,  there is no known example of natural  cellular processes that  extract work from correlations. Instead, when no longer needed,   biomolecules are typically actively broken down at a cost to the cell. This suggests that the  low entropy  contained in those molecules is not exploitable. 
At the same time, however, there are examples in biological systems that exhibit behavior similar to that of a random access symbol gas. In particular ATP/ADP stores, whilst behaving primarily as a store of chemical energy, also exert an entropic force in the manner of an information reservoir.
\par


\section{Model \& Work calculation.}

\begin{center}
\begin{figure*}
\includegraphics[width=0.75\textwidth]{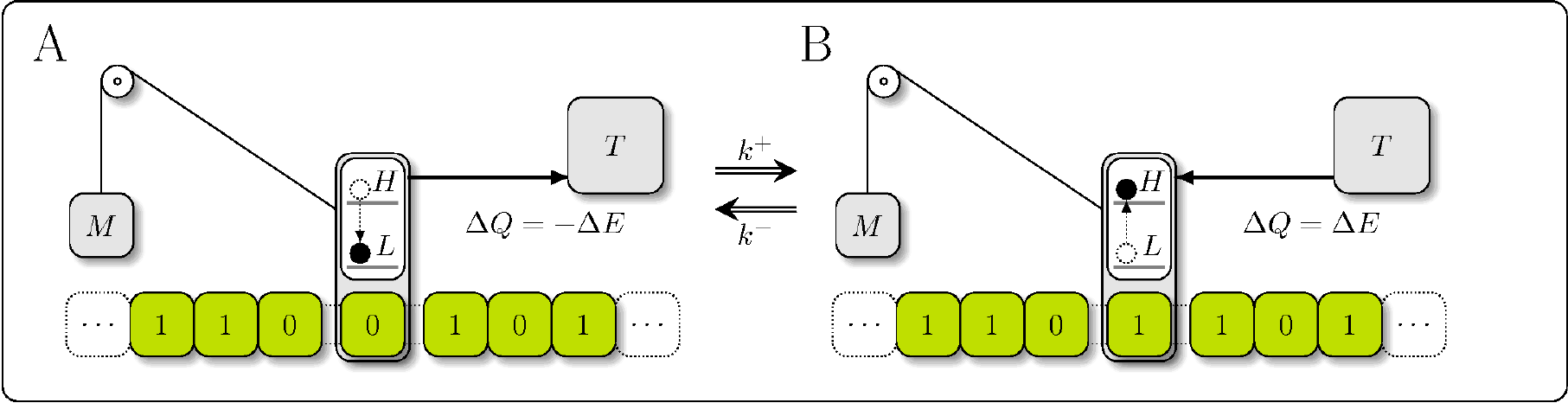}
\includegraphics[width=0.75\textwidth]{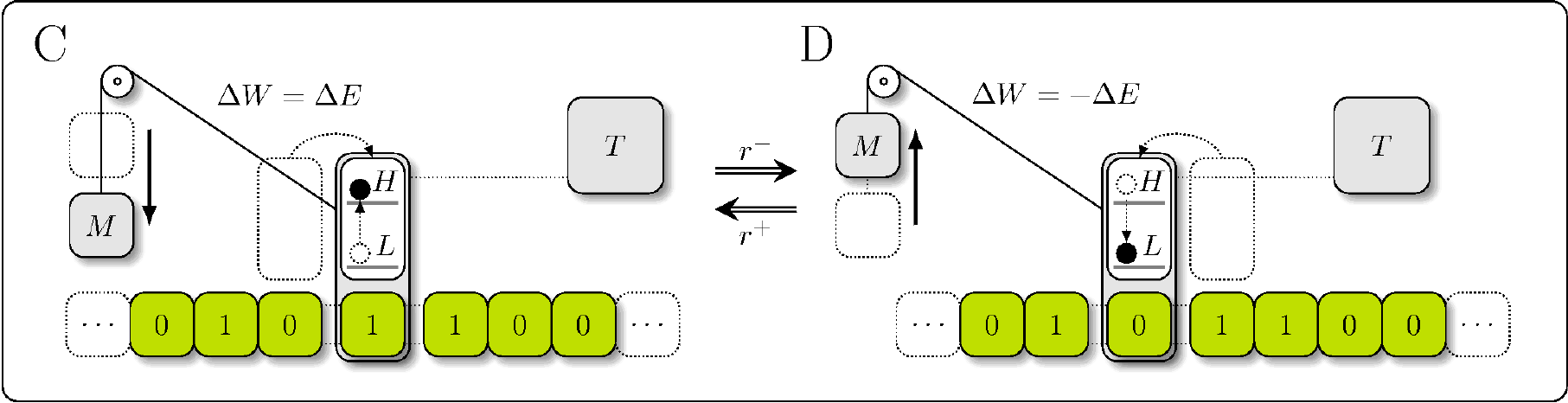}

\caption{Operation of the demon: Four instances of the demon interacting with a work reservoir (drawn as a mass on a pulley), a heat reservoir at temperature $T$ and an information reservoir (drawn as a tape)  and transitions between them. Indicated work and heat flows are defined as on and to the system respectively.  
The transition from instance A to B is a thermal excitation of the demon (with rate $k^+$) accompanied by overwriting the reservoir site with a $1$. The transition from instance B to A is a thermal relaxation of the demon (with rate $k^-$) accompanied by overwriting the reservoir site with a $0$. The transition from instance C to D is a spatial transition to the left (with rate $r^-$) resulting in the bound reservoir symbol changing from a $1$ to a $0$ whereby energy is captured by the work reservoir. The transition from instance D to C is a spatial transition to the right (with rate $r^+$), causing the bound reservoir symbol to change from a $0$ to a $1$, whereby work is expended by the work reservoir.
\label{demonabc}
}
\end{figure*}
\end{center}

We introduce a demon based on that described in \cite{barato_stochastic_2014}. This is a device which is characterised by two states  $\{H,L\}$ with energies $\Delta E$ and $0$ respectively. This device is coupled to both a heat bath at temperature $T$ and a work reservoir. At any one time the demon is also in contact with, or bound to, a single element of an information reservoir, see Fig.~(\ref{demonabc}). This information reservoir consists of  symbols in the alphabet $\{0,1\}$. No energy differences exist between these two symbols. We assume that the  information reservoir is intialised such that, for each element, the symbol $1$ occurs with probability $\epsilon$. We further assume that the states of the demon and the symbols of the reservoir correspond to one another. When  bound to a given reservoir element  the demon is  always in the  $\{H\}$ state when the current reservoir element is in state $1$. Correspondingly,   the demon is in the  $\{L\}$ state  when the current reservoir element is  $0$. Consequently, thermal fluctuations caused by exposure to the heat bath induce the joint transitions $\{H,1\}\leftrightarrow\{L,0\}$ such that the symbol associated with the currently bound element of the information reservoir is overwritten. These fluctuations are characterised by excitations, $\{L,0\}\rightarrow\{H,1\}$, occurring with rate $k^+$ and relaxations, $\{H,1\}\rightarrow\{L,0\}$, occurring with rate $k^-$, related by $k^-/k^+=\exp(\Delta E/T)$, which are accompanied by an exchange of heat with the heat bath. (Note we set $k_B=1$ here and throughout.) The timescale of such fluctuations is controlled by $k=k^++k^-$. However, to operate upon the information reservoir as a whole, and extract work from the heat bath, the demon must change the reservoir element to which it is bound.
\par
 It is here, and in contrast to the BS model \cite{barato_stochastic_2014}, that we insist that the information reservoir has a physical structure. Specifically, we interpret the  elements of the information reservoir  as \emph{sites} on a lattice. For instance, the reservoir might be one dimensional, i.e. a tape.
\par
We now consider the ability of the demon to access distinct sites in such a physically structured reservoir. Consequently, we allow the demon to undergo spatial transitions at some rate $\gamma$ between neighbouring sites. When this occurs the symbol associated with the reservoir site bound to the demon may change. If this happens the demon is required to change state, $\{H\}\leftrightarrow\{L\}$, also, with an associated expense, or recovery, of energy either provided by, or stored in, the work reservoir. Such behaviour is illustrated in the lower panel in Fig.~(\ref{demonabc}). The transition from C to D corresponds to a mechanical relaxation of the demon due to the intervention of the work reservoir as the bound symbol changes from a $1$ to a $0$ as the demon changes reservoir site. Explicitly, in this case, the demon moves one site to the left on the reservoir, here pictured as a tape, changing the symbol bound to the demon from a $1$ to a $0$. This is necessarily accompanied by a drop from state $\{H\}$ to state $\{L\}$ in the demon, providing an excess $\Delta E$ of energy which is stored in the work reservoir. Similarly, the transition from D to C corresponds to a mechanical excitation of the demon due to the intervention of the work reservoir as the bound symbol changes from a $0$ to a $1$. Explicitly, the demon moves one site to the right on the tape, changing the symbol bound to the demon from a $0$ to a $1$. This is necessarily accompanied by a raising the  state of the demon from  $\{L\}$ to $\{H\}$, requiring $\Delta E$ energy which is provided by the work reservoir. Transitions  to new reservoir sites where  the symbol does not change  incur no such exchange of work; see Fig.~(\ref{demonabc}) for a summary of the basic transitions that incur energy exchange.
\par
To model this system we now assume that the demon performs a biased random walk on the information reservoir in order to reach new sites. In 1D it steps to the right with rate $r^+$ and steps to the left with rate $r^-$ such that the total rate of spatial transitions is $\gamma=r^++r^-$. Because of this random walk the demon may, however, transition to sites that have been visited previously.
\par
Throughout this contribution we  assume a separation of time-scales, $\gamma \ll k$, which implies that the currently bound reservoir site is equilibrated before the demon moves to a neighbouring site. It thus extracts any informational resource during the first interaction with the site.  This entails that   sites that have been previously visited have symbol 1 with probability $p_s$, equal to the equilibrium probability of the demon being in state $\{H\}$.  The statistics of the unvisited sites is different. These, have symbol 1 with probability $\epsilon$.
\par
This allows us to formulate a mean-field model where  all spatial degrees of freedom associated with the information reservoir are replaced by a single degree of freedom concerning whether or not the currently bound site has been previously bound to the demon. Each step of the demon thus amounts to a random draw of sites, where a site can be either previously visited or unvisited and either contain symbol  0 or  symbol  1, each with a time-dependent probability. The demon thus has 4 pseudo-states $\{\{H,\text{new}\},\{H,\text{old}\},\{L,\text{new}\},\{L,\text{old}\}\}$. The labels `$\text{old}$' and `$\text{new}$' indicate that the currently bound site has and has not been previously bound to the demon, respectively.  
\par
\begin{figure}
\includegraphics[width=0.47\textwidth]{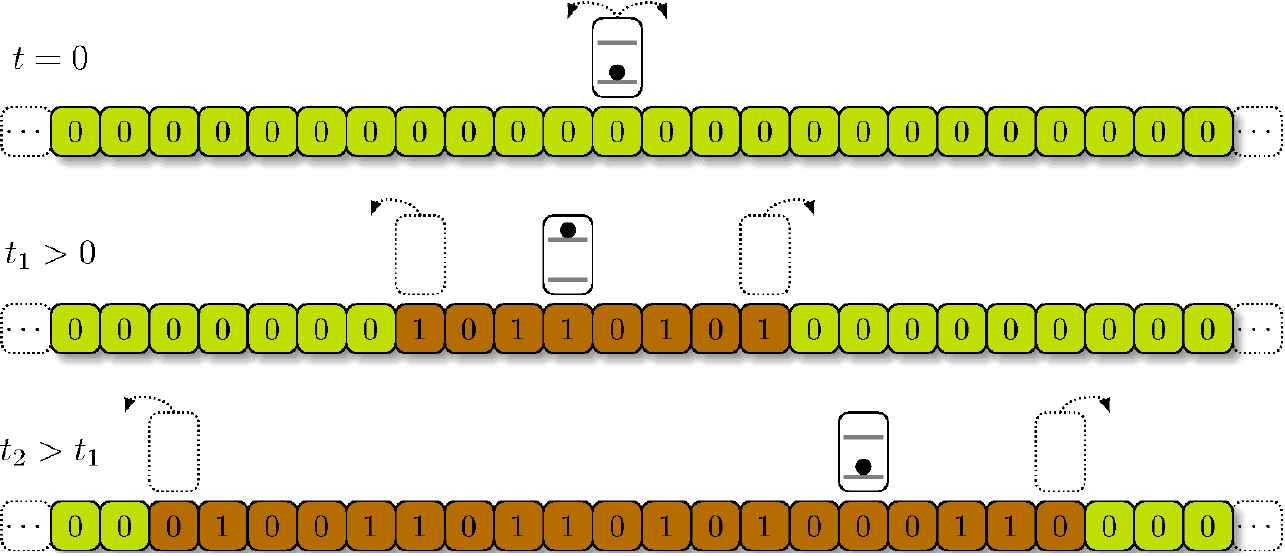}
\caption{Visited and unvisited reservoir sites: Shown is a demon exploring a 1D information reservoir (tape) at different times since initialisation. Previously unvisited sites are lightly shaded with statistics $\epsilon =0$ (all $0$s). Darker sites are previously visited sites with different statistics, given by $p_s$. At time $t=0$ all sites are previously unvisited and all transitions will be to an unvisited site. At subsequent times, dotted outlines indicate the location and transitions required to arrive at a previously unvisited site. At time $t_2>t_1$ more sites have been visited, but still only two transition/location combinations will result in becoming bound to an unvisited site. This causes $P_u(d,p,t)$ to reduce over time. }
\label{newsites}
\end{figure}
\begin{figure}
\includegraphics[width=0.47\textwidth]{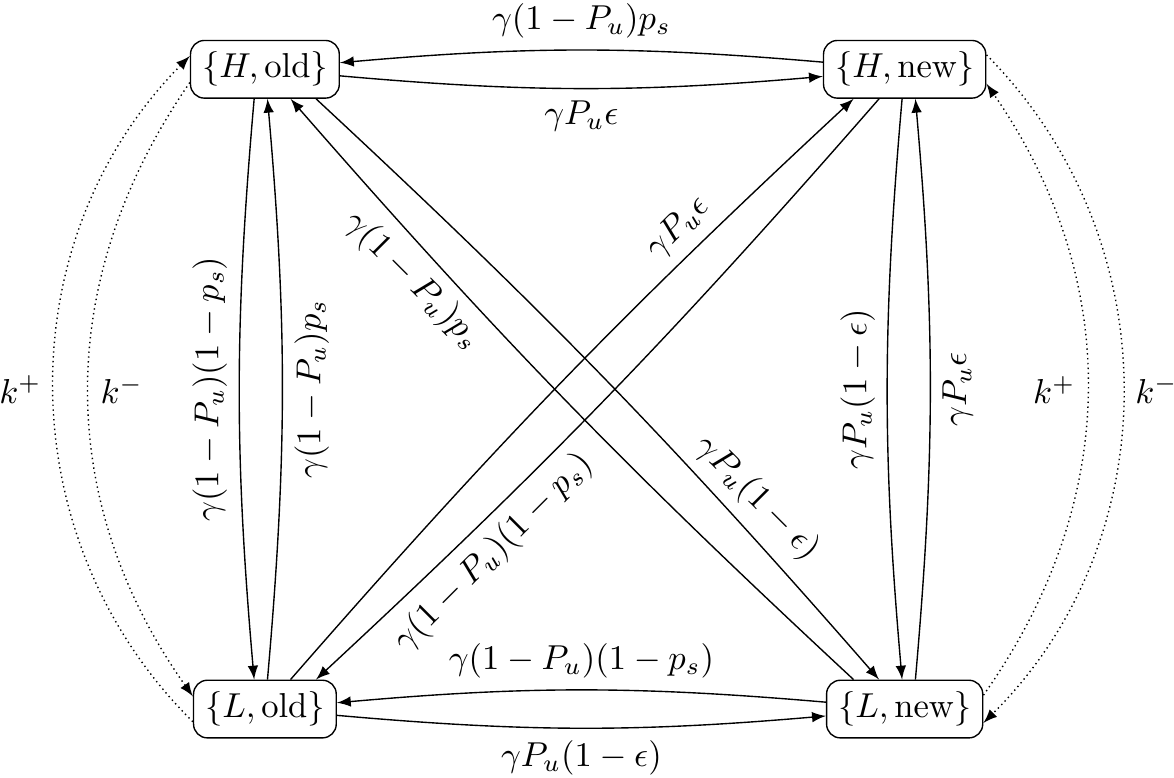}
\caption{Transition diagram for the spatial demon. The transition rates relating to exploration of the information reservoir are indicated on the transition diagram  with solid arrows. Additionally, there are thermal transitions with rates $k^-$ and $k^+$  indicated with dotted arrows.}
\label{transitions}
\end{figure}

A crucial function for the  remainder of this contribution is the function  $P_u(d,p,t)$, which denotes  the probability that any given transition corresponding to exploration of the information reservoir is to a previously unvisited site. This probability is derived from the random walk statistics which is dependent on the structure of the reservoir and the dynamics of the walker. It depends on the bias of the walk, $p$, the dimension of the tape, $d$, and the time $t$ since the demon first came into contact with the tape (see Fig.~(\ref{newsites})). The model then consists of various transitions between the four pseudo-states, the rates of which are detailed in Fig.~(\ref{transitions}).
\par
The  transitions indicated in Fig.~(\ref{transitions}) are sufficient to formulate a master equation which, in the limit of timescale separation $\gamma\ll k$, yields the steady state probability for the demon to be in state $\{H\}$ as simply the thermal equilibrium distribution
\begin{align}
p_s&=P(\{H,\text{old}\})+P(\{H,\text{new}\})=\frac{k^+}{k^++k^-}.
\end{align}
We now calculate the average work extracted  when transitioning to  new, unvisited, sites:
\begin{align}
&\langle\dot{W}_{\text{new}}\rangle\nonumber\\
&= \left(p_s\; w(\to \{L,\text{new}\})\:\:-(1-p_s)\; w(\to \{H,\text{new}\})\right)\Delta E\nonumber\\
&=(p_s\gamma P_u(d,p,t)(1-\epsilon)-(1-p_s)\gamma P_u(d,p,t) \epsilon)\Delta E\nonumber\\
&=\gamma P_u(d,p,t)(p_s-\epsilon)\Delta E.
\end{align}
Here, $w(\to \{L,\text{new}\})$ is the transition rate to the state  $\{L,\text{new}\}$ from the states $P(\{H,\text{old}\})$ and $P(\{H,\text{new}\})$ and can be read off from Fig.~(\ref{transitions}).  The work extracted from the visited tape elements is analogously given by
\begin{align}
&\langle\dot{W}_{\text{old}}\rangle\nonumber\\
&= \left(p_s\; w(\to \{L,\text{old}\})\:\:-(1-p_s)\; w(\to \{H,\text{old}\})\right)\Delta E\nonumber\\
&=(p_s\gamma P_u(d,p,t)(1-p_s)-(1-p_s)\gamma P_u(d,p,t) p_s)\Delta E\nonumber\\
&=0,
\end{align}
which vanishes since the the distributions characterizing the demon and visited reservoir sites are identical. 
The mean work extracted per tape transition, $\langle\dot{W}_{\text{ext}}\rangle=\langle\dot{W}_{\text{new}}\rangle+\langle\dot{W}_{\text{old}}\rangle$, then reads
\begin{align}
\gamma^{-1}\langle\dot{W}_{\text{ext}}\rangle &=\frac{\left(k^+-\epsilon(k^++k^-)\right)P_u(d,p,t)) T}{k^++k^-}\ln\left(\frac{k^-}{k^+}\right)\nonumber\\
&=\frac{(1-\epsilon(1+e^{\frac{\Delta E}{T}}))\Delta E P_u(d,p,t)}{(1+e^{\frac{\Delta E}{T}})}. \label{work}
\end{align}
As in the MJ model, the device can operate in three phases characterized by \one positive work extraction, \two a reduction in entropy of the tape (an information eraser) and \three no extraction or erasure, a `dud'. In this contribution we discuss only work extraction, but note similar arguments for information erasure follow.
\par
An {\em optimal demon} maximizes   work extraction, which is achieved by  choosing $\epsilon=0$ and $\Delta E=T(1+\mathcal{W}(e^{-1}))$  (shown in Appendix \ref{smIC}) such that 
\begin{align}
\gamma^{-1}\langle\dot W_{\text{max}}\rangle&=\mathcal{W}(e^{-1})P_u(d,p,t) T\nonumber\\
&\simeq 0.27P_u(d,p,t) T
\end{align}
 where $\mathcal{W}$ is the Lambert W function. This is the maximum work extraction rate using an information reservoir, but note that this excludes any cost associated with operation of the demon. In our model operational costs arise due the biasing of the random walk performed by the demon that enables exploration of the information reservoir.

\subsection{Cost of operation}

 In order to obtain an understanding of the net-amount of work that can be rectified from the heat reservoir in conjunction with the information reservoir, the cost of driven exploration of the information reservoir needs to be accounted for.  In the MJ model such a cost was not considered, whilst in the BS model no such cost was incurred because their model permitted random access to sites due to the implicit assumption that the information reservoir was organized as a perfectly mixed gas; hence there was no spatial exploration to drive.
\par
To understand the costs of driven exploration of an information reservoir, we first consider a 1D structure, i.e. a tape.  As we assume that transition to the right  occur with rate $r^+$ and to the left with rate $r^-$ and require $\gamma=r^++r^-$, in characteristic time $\gamma^{-1}$ we can expect a transition to the right with probability $p=r^+/(r^++r^-)$, which characterises the bias of the walk. An unbiased walk corresponds to $r^+=r^-$ and thus $p=1/2$. Any biasing, however, comes at the expense of energy input, in the form of work. Consequently, assuming exposure to a heat bath at temperature $T$, and a local detailed balance relation \cite{bergmann_new_1955,bauer_local_2015}, the work rate required to explore the (1D) reservoir (per transition) is given by
\begin{align}
\gamma^{-1}\langle \dot{W}_{\text{tape}}\rangle&=T\frac{r^+}{r^++r_-}\ln\left(\frac{r^+}{r^-}\right)+ T\frac{r^-}{r^++r^-}\ln\left(\frac{r^-}{r^+}\right)\nonumber\\
&=T (2p-1)\ln\frac{p}{1-p}, \label{tape}
\end{align}
where the transition probability along the tape is independent of the energetics of the demon/tape coupling with the changes in energy achieved through the intervention of the work reservoir. Importantly, for the optimal demon, both $\langle \dot{W}_{\text{max}}\rangle$ and $\langle \dot{W}_{\text{tape}}\rangle$ have a simple linear dependence in the temperature $T$ allowing for temperature independent results, provided the thermal and spatial transitions of the demon originate in heat baths at the same temperature.

\section{Exact transient solution for a demon and 1D tape.}
\label{oned}

Here, we utilise the work calculations detailed above and investigate the net performance of such a demon operating on a 1D information reservoir by computing an exact expression for the probability to encounter an unvisited site at time $t$ under a bias $p$, $P_u(1,p,t)$.
\par 
 To do so we first consider the probability that a discrete time random walker will arrive at an unvisited site on its $n$-th transition, $P_u(1,p,n)$. A complete derivation is given in Appendix \ref{smI}, but the structure of the derivation is as follows. We identify the probability of arriving at an unvisited site on the $n$-th transition as the change in the expected number of visited sites, $S(n)$, between times $n$ and $n-1$. By considering the generating functions of first passage probabilities and their relation to the generating function of the probability of general site occupancy of a biased walker \cite{montroll_random_1965} we find the generating function of $S(n)$ to be
 \begin{align}
 S(z)&=\frac{(1+4zp(1-p))^\frac{1}{2}}{(1-z)^2}.
 \end{align}
 The probability of visiting a new site at time $n$, $S(n)-S(n-1)$, is then derived by considering differences between standard usages of the generating function
 \begin{align}
 S(n)&=\frac{1}{n!}\frac{\partial^n}{\partial z^n}S(z)\Big|_{z=0}.
 \end{align}
 Further manipulation then allows us to write:

  \begin{align}
 &P_u(1,p,n) =|1{-}2p|\nonumber\\
 &+\frac{(4p(1-p))^{m}\Gamma\left[m{-}\frac{1}{2}\right] {}_2F_1\left(1,m{-}\frac{1}{2};m{+}1;4p(1{-}p)\right)}{2\sqrt{\pi}\Gamma\left[m+1\right]}
 \end{align}
where $_2F_1$ is the hypergeometric function and $m=\ceil*{(n+1)/2}$. This gives the long term limit $\lim_{n\to\infty}P_u(1,p,n)=P^{s}_u(1,p)=|1-2p|$ independently of the time basis. For $p=0,1$ this probability is always $1$, whilst for $p=1/2$ in the $n\to\infty$ limit the probability vanishes. Intuition can be gained from Fig.~(\ref{newsites}): for an unbiased, diffusive walker, the probability of being at the edge of the domain of visited sites eventually vanishes, but an entirely biased walker is always at the edge of such a domain and steps away from the visited sites. In particular, for $p=1/2$ we have
\begin{align}
P_u(1,1/2,n)=\frac{\Gamma[m-1/2]}{\sqrt{\pi}\Gamma[m]}.
\end{align}
Conversion to a (Markov) continuous time process gives 
\begin{align}
P_u(1,p,t)=\sum_{n=0}^{\infty}P_u(1,p,n)\frac{e^{-\gamma t}(\gamma t)^n}{n!}
\end{align}
 which leads to an exact form for the work extraction rate by substituting into Eq.~(\ref{work}),  provided $\gamma\ll k$ holds.  An asymptotically exact solution exists for $p=1/2$, $P_u(1,1/2,t)\simeq (2/(\pi\gamma t))^{1/2}$.  Convergence to the $t\to \infty$, steady state, behavior consequently follows a power law for $p=1/2$, but exhibits increasingly fast exponential tails as $p$ deviates to $0,1$.  A numerical example can provide some intuition for the behavior of $P_u$ over time. In the case of no-bias --- $p=1/2$ --- the probability of discovering a new site has dropped to around $0.025$ and $0.008$ after  $\sim 1000\gamma^{-1}$ and $\sim 10000\gamma^{-1}$ seconds respectively.  Consequently, the unbiased walker experiences a steady reduction of its work extraction rate as its operation time increases, illustrated further in Appendix \ref{smI}.
\par
For the remainder of this paper we shall be concerned exclusively with the steady state net-extraction rate of work,  given by the \emph{difference}
\begin{equation}
\gamma^{-1}\langle\dot{W}_{\text{net}}\rangle :=\gamma^{-1}\langle\dot{W}_{\text{ext}}\rangle-\gamma^{-1}\langle\dot{W}_{\text{tape}}\rangle.
\end{equation}
We now come to the key insight of this present contribution: A fair assessment of the performance of any such demon  must take into account  the statistics of $\langle\dot{W}_{\text{net}}\rangle$ rather than $\langle\dot{W}_{\text{ext}}\rangle$, which has normally been considered. And, importantly, both terms that form this difference, $\langle\dot{W}_{\text{ext}}\rangle$ and $\langle\dot{W}_{\text{tape}}\rangle$, are dependent on the bias, $p$. For instance, in the undriven regime ($p=1/2$) both the work extracted  and the work required  to bias the exploration of the information reservoir are $0$ --- a purely diffusive tape requires no work to drive it, but equally the probability to find new sites $P_u\to 0$ goes to zero for the unbiased walker, hence no work extraction is possible.   
\par
In order to gain insight into  the work extraction rates possible for biased explorations of tapes ($p\neq 1/2$), we expand  the work extracted (Eq.~\ref{work}) and work expended (Eq.~\ref{tape}) around  $p=1/2$. This analysis yields a  {\em linear} dependence in $(p-1/2)$ in the extracted work, but {\em quadratic} dependence in the work spent driving the tape
\begin{align}
\gamma^{-1}\langle\dot {W}_\textrm{tape}\rangle&\simeq 8T\left(p-{1}/{2}\right)^2+\mathcal{O}\left(\left(p-{1}/{2}\right)^3\right).
\label{tape_expand}
\end{align}
This result implies that it is \emph{always} possible to extract net positive work from the 1D tape in the limit of small (but non-vanishing) bias, i.e. $|p-1/2|\gtrsim 0$, provided that  the random access model is capable of work extraction.  
\par
However, as the bias is increased further, the work required to explore the information reservoir rises faster than the resultant increase in the incoming rate of unvisited sites. There is thus a critical bias $p_c>1/2$, beyond which no net-extraction of work is possible. We find this critical bias to be given by (for details see Appendix \ref{smIC}) 
\begin{equation}
p_c=(1+e\mathcal{W}(e^{-1}))^{-1}\simeq 0.569.
\end{equation}
 Furthermore for a 1D reservoir, there is an optimal bias $p_{\text{opt}}$ where the net-extraction of work is optimized;  we find that it fulfils  (see Appendix \ref{smIC} for details) 
\begin{equation}
p_{\text{opt}}=(1-p_{\text{opt}})\exp[{\mathcal{W}(e^{-1})+\frac{1}{2}\left(p_{\text{opt}}^{-1}-(1-p_{\text{opt}})^{-1}\right)}].
\end{equation}
The value of $p_{\text{opt}}$ needs to be determined numerically. We  found it to be a rather modest bias of $p_{\text{opt}}\sim 0.535$.
\par
 Next, we ask what impact the spatial structure of the information reservoir has on the performance of the demon.  To understand this, we  use  the performance of the BS demon,  that  always samples unvisited reservoir sites with zero expenditure of work as a reference. We thus define the efficiency parameter   $\eta$ as  the ratio of the net work extraction of an optimally set up demon, with respect to model parameters and bias $p_{\rm opt}$, to the total work that can be extracted from an optimally set up BS demon. Note that the work extraction of the BS demon  is the work extraction of our model in the case when the demon is perfectly biased, i.e.  $p=1$ (or $p=0$) in our model.  As such this reference work performance, in the $\gamma\ll k$ regime, is given by $\gamma^{-1}\langle\dot{W}_{\text{max}}\rangle_{p=1} =\mathcal{W}(e^{-1})T\simeq 0.27 T$.  In Appendix \ref{smIC} we find the efficiency parameter to be given by 
\begin{align}
\eta&=(2p_{\text{opt}}-1)\left(1+\mathcal{W}(e^{-1})^{-1}\ln\frac{p_{\text{opt}}}{1-p_{\text{opt}}}\right)\nonumber\\
&\simeq 0.035.
\end{align}
 This shows that the  performance of a demon operating on a 1D tape is dramatically reduced. 
\par
\begin{figure}
\includegraphics[width=0.2\textwidth]{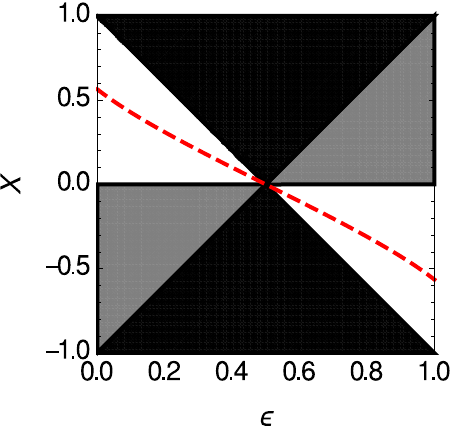}
\includegraphics[width=0.2\textwidth]{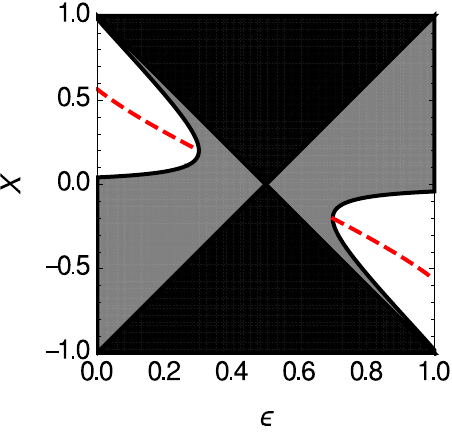}
\includegraphics[width=0.2\textwidth]{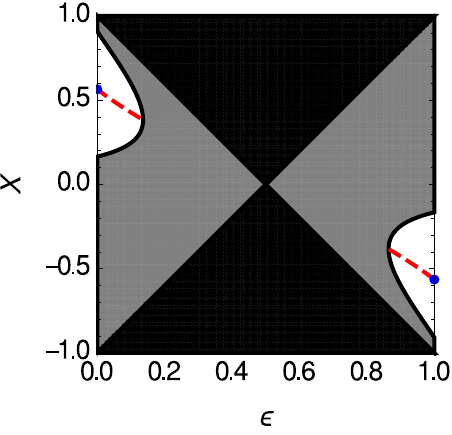}
\includegraphics[width=0.2\textwidth]{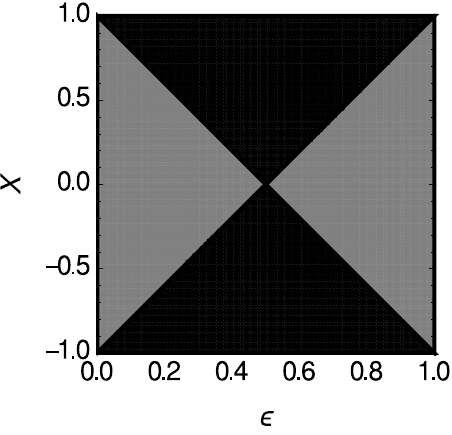}
\caption{Phase diagram in 1D. Black indicates an information erasing phase, white areas work extraction $\langle\dot{W}_{\text{net}}\rangle>0$, and gray areas `dud' phases where there is no information erasure and $\langle\dot{W}_{\text{net}}\rangle<0$. The $x$ axis is $\epsilon$, whilst the $y$ axis is $\mathcal{X}:=(e^{\Delta E/T}-1)/(e^{\Delta E/T}+1)$. From top left to bottom right we utilize $p=(1/2)^+$, $p=0.51$, $p=p_{\text{opt}}\simeq0.5348$ and $p>p_c\simeq 0.569$ respectively. Red lines indicate contour of optimal $\Delta E$ given $\epsilon$. Blue points indicate coordinates of maximal performance. We set  $T=1$ and $k=k^++k^-=1$.}
\label{phase}
\end{figure}

  Finally, we note that the consideration of net extracted work not only has an effect upon the maximum possible performance of such a demon, but also the phase diagram of its operation. By considering the net work balance we must come to redefine the phases such that the work extracting phase corresponds to sets of parameter values where $\langle\dot{W}_{\text{net}}\rangle>0$ and the `dud' phases correspond to sets where $\langle\dot{W}_{\text{net}}\rangle<0$ and the entropy of the reservoir is being increased ($|p_s-1/2|<|\epsilon-1/2|$). The consequence of doing so has a dramatic effect and is illustrated in Fig.~(\ref{phase}). Only in the limit $p\to 1/2$ (where $\langle \dot{W}_{\rm tape}\rangle$ is negligible compared to $\langle \dot{W}_{\rm ext}\rangle$) does the phase diagram replicate that of the BS and MJ models, but, of course, the net work extraction rate here approaches $0$.  For higher values of $p$ the combinations of demon parameters that will successfully lead to net work extraction is significantly reduced, until for $p\geq p_{c}$ no net work extraction is possible at all.

\section{Generalization to $d>1$ information reservoirs.}

We now generalize from 1D tapes to information reservoirs with $d>1$, thus bridging the gap between the 1D MJ and 
random access BS models, which we associate with spatially structured reservoirs of infinite dimension. 
We can characterize such situations by considering a demon that performs a random walk on arbitrary $d$-dimensional simple cubic lattices, $\mathbb{Z}^d$, for which, as before, each site is in a state corresponding to  the symbol $0$ or $1$. If the rates governing exploration of the reservoir in each positive and negative spatial dimension are $r^+/d$ and $r^-/d$ then both the timescale, $\gamma$, and the work rate associated with driving such a random walk  is independent of the dimensionality. To proceed, we will first derive an expression for the steady-state probability  $P^{s}_u(d,p)$  that a random walker in $d$ dimensions with a bias $p$ will encounter a previously unvisited lattice site in its next transition. Mirroring section \ref{oned}, we will then use $P^{s}_u(d,p)$  in order to understand the scope for work extraction using higher dimensional information reservoirs.  %
 \begin{figure}
\includegraphics[width=0.45\textwidth]{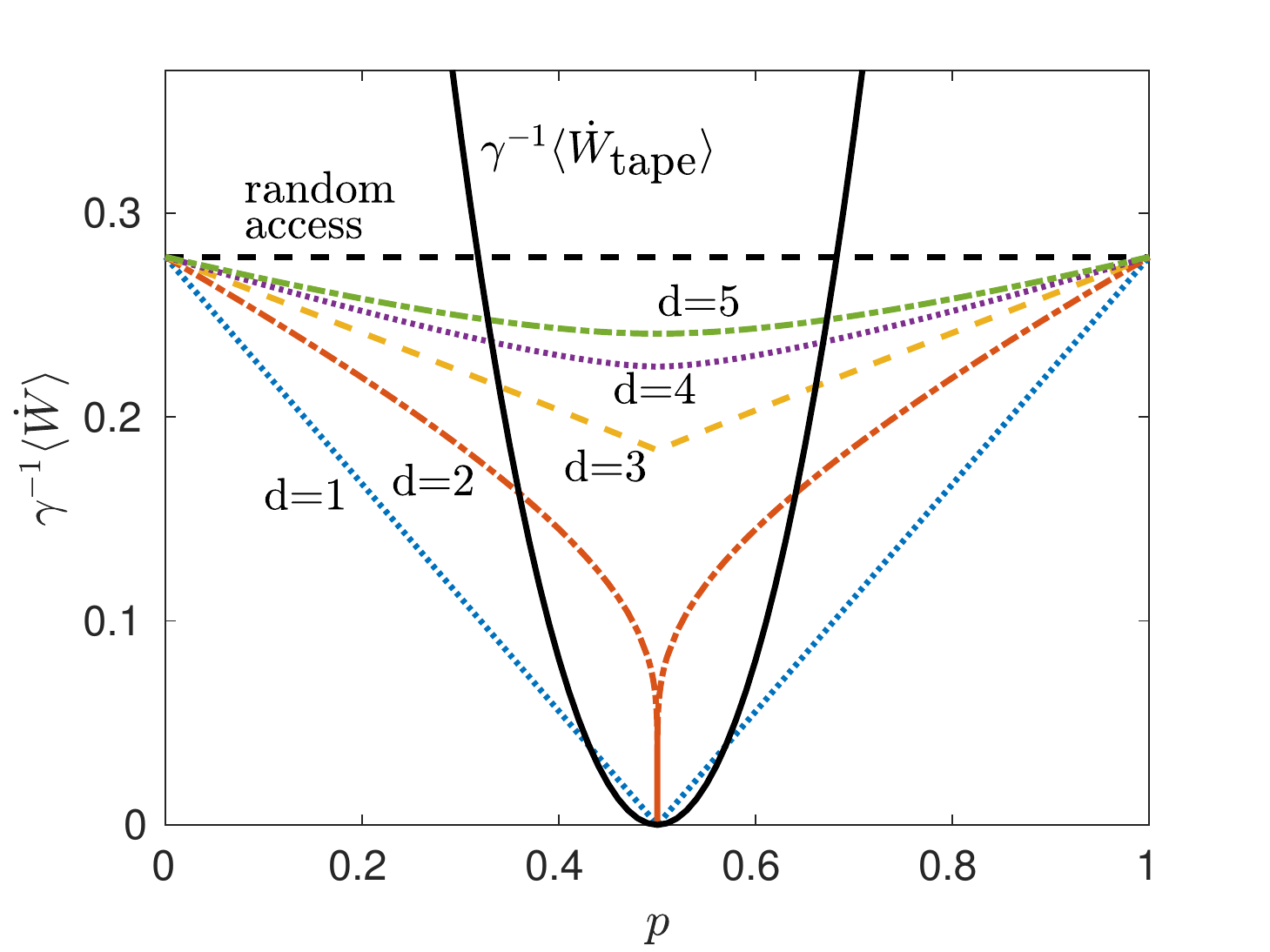}
\caption{Multi-dimensional information reservoirs. $\langle\dot{W}_{\text{tape}}\rangle$ is in black. $\langle\dot{W}_{\text{ext}}\rangle$ in dimensions 1-5 is also shown. Finally the dashed line indicates the $n\to\infty$, random access, limit. $\langle\dot{W}_{\text{net}}\rangle>0$ occurs when the extraction curves lie above the cost of driving. This is maximized for given $p$. We use $\Delta E=T(1+\mathcal{W}(e^{-1}))$, $T=1$ and $\epsilon=0$.}
\label{multikulti}
\end{figure}
\par
In order to derive $P^{s}_u(d,p)$ we  adapt the derivation given by    Montroll \cite{montroll_random_1956}  to  compute  the probability of recurrence for  a random walker subject to a uniform bias in all dimensions.  Full details can be found in Appendix \ref{smII}, but we provide the salient aspects here. 
\par
We concern ourselves firstly with the probability, $P(\mathbf{x},n|\mathbf{o},0)$, of being at a lattice site $\mathbf{x}=\{x_1,\ldots,x_d\}$, with dimensionality $d$, at time $n$, having started at the origin $\mathbf{o}$ at time $0$. We may then consider the following generating function 
\begin{align}
G(\mathbf{x},z)&=\sum_{n=0}^{\infty}P(\mathbf{x},n|\mathbf{o},0)z^{n}.
\end{align}
By considering a system with periodic boundary conditions and taking the infinite size limit, this generating function may be expressed in the general form
\begin{align}
G(\mathbf{x},z)&=\frac{1}{(2\pi)^d}\underbrace{\int_0^{2\pi}\ldots \int_0^{2\pi}}_{\text{d times}}\frac{\exp[i\theta\cdot\mathbf{x}]}{1-z\lambda(\theta)}d\theta_1\ldots d\theta_d,
\end{align}
where $\theta=(\theta_1,\ldots,\theta_d)$ and $\lambda(\theta)$ is a so-called structure function 
\begin{align}
\lambda(\theta)=\sum_{\mathbf{x}}P(\mathbf{x},n+1|\mathbf{x}',n)\exp[i(\mathbf{x}-\mathbf{x}')\cdot\theta]
\end{align}
which, here, encodes the one step probabilities of the random walker. For the $d$-dimensional simple cubic lattices we are considering, the structure function is of the form
\begin{align}
\lambda(\theta)=d^{-1}\sum_{i=1}^d p\exp[i\epsilon_i\theta_i]+(1-p)\exp[-i\epsilon_i\theta_i]
\end{align}
where $p\in[0,1]$ is the bias parameter, with $p=1/2$ corresponding to no bias, and $\epsilon_i\in\{1,-1\}$ specifies the direction of the bias in dimension $i$. This allows computation of the generating function. It is possible to relate such a generating function to the generating function concerning \emph{first} arrival at a given site which, in turn, allows an expression for the probability of \emph{ever} encountering such a site through the consideration of $1-(G(\mathbf{x},1))^{-1}$ \cite{montroll_random_1956}. Setting $\mathbf{x}=\mathbf{o}$ allows us to compute the probability of eventual recurrence to the origin, $P_r(d,p)$. This is related to the rate of encountering new sites in the long time limit by $1-P_r(d,p)=\lim_{n\to \infty}dS(n)/dn$ \cite{a._dvoretzky_problems_1951,vineyard_number_1963}, where, again $S(n)$ is the expected number of visited sites. Further manipulation allows us to find the following expression for such a probability in terms of the dimensionality of the lattice and bias of the walker along the axes (importantly, noting independence in the $\epsilon_i$ terms controlling the direction of the bias)

\begin{align}
P_u^s(d,p)&=\left[\int_0^\infty e^{-x}\left[I_0\left(\frac{2x}{d}\sqrt{p(1-p)}\right)\right]^d dx\right]^{-1},
\label{int2}
\end{align}
where $I_0$ is the zero-th modified Bessel function of the first kind. 
We note that for all $d$, if $p=0$ or $p=1$ then the integral reduces to $\int_0^{\infty} e^{-x}dx=1$ indicating an escape probability, and thus $P_u^s(d,p)$, of $1$ at total bias (complete irreversibility). For $d=1$ and $d=2$ Eq.~(\ref{int2})  can be evaluated exactly, thus recovering the 1D result $P^{s}_u(1,p)=|1-2p|$, and the 2-D result $P^{s}_u(2,p)=\pi/(2K[4p(1-p)])$ where $K$ is the complete elliptic integral of the first kind. It should be noted that Eq.~(\ref{int2}) concerns only simple cubic lattices, however, analogous forms for body-centered cubic and face-centered cubic structures exist, related to the well known Watson integrals. For details on the extension of such lattices to $d$ dimensions and triangular lattices in 2D, with the addition of biasing, see Appendix \ref{AppB2}. The formulae for recurrence on these lattices, however, cannot be expressed in a way that makes further analysis possible and so we proceed with simple cubic structures and Eq.~(\ref{int2}) only. 
\par
In full analogy to section \ref{oned}  we can now use Eq.~(\ref{int2}) to calculate the net-amount of extracted work numerically.  The resulting extracted work for an optimal demon operating on simple cubic reservoirs of dimension $1$ through $5$ for varying bias, and the corresponding cost of exploring the reservoir at that bias, are shown in Fig.~(\ref{multikulti}). We find that as the dimension increases, the work extracted  also increases, approaching that of the  limiting case of the BS model. This implies that net-extraction of work by the demon becomes increasingly efficient  as the dimension of the information reservoir increases,  reproducing the  random access model in the $d\to\infty$ limit as expected. 
\par
We note that Eq.~(\ref{int2}) has continued the domain to $d\in\mathbb{R}_+$, where it remains analytic for $p\neq 1/2$. We discuss the behavior of such a function with $d\in\mathbb{R}_+$ in order to understand better  the case $d\in\mathbb{Z}_+$, but remain agnostic on its physical significance.  
\par
\begin{figure}
\begin{centering}
\includegraphics[width=0.47\textwidth]{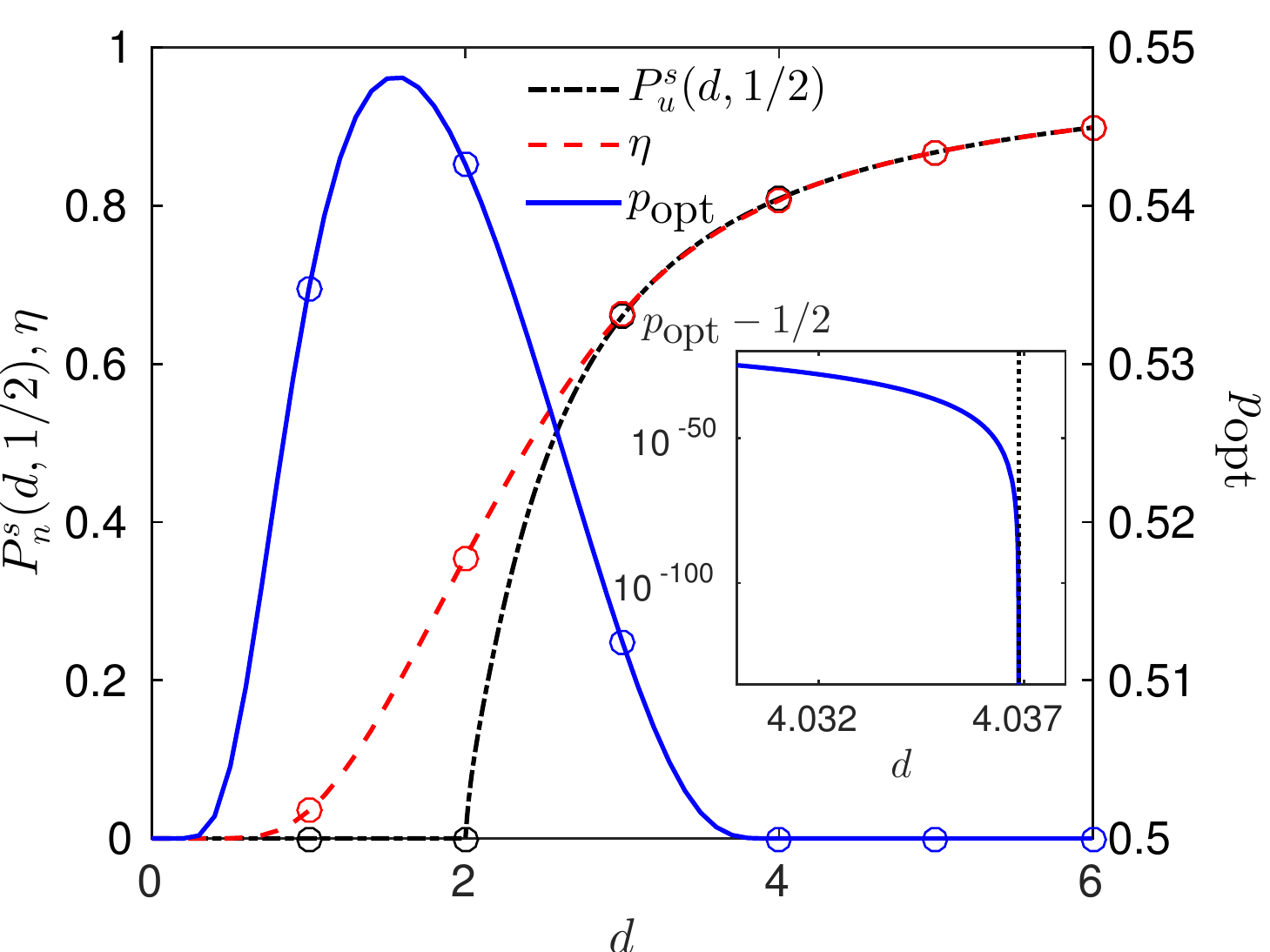}
\caption{Behavior of $P^{s}_u(d,1/2)$, $p_{\text{opt}}\geq 0.5$ and $\eta$ varying with $d$ with integer $d$ marked. $P^{s}_u(d,1/2)$ separates from the axis at the critical value of $d=2^+$, leading to distinct behaviour for integer dimensions $d\leq 2$ and $d\geq 3$. The approach, $p_{\text{opt}}\to 1/2$ at critical value $d\sim 4.0369$ is shown inset, leading to distinct optimal behaviour for integer dimensions $d\leq 4$ and $d\geq 5$.}
\label{d_crit}
\end{centering}
\end{figure}

 Firstly, we note that $P_u^s(d,1/2)=0$ for $d\leq 2$, reflected in the specific results $P^s_u(1,1/2)$ and $P^s_u(2,1/2)$. However, for $d> 2$ $P_u^s(d,1/2)$ is positive.  Based on this we identify  a critical value of $d=2^+$, separating the integer dimensions $d\leq 2$ and $d>2$.  When $d>2$ work can be extracted in the limit $t\to\infty$ at $0$ bias and below this value ($d\leq 2$) no work can be extracted.  This behavior, continued into $d\in\mathbb{R}_+$, is shown in Fig.~(\ref{d_crit}), but can also be observed in Fig.~(\ref{multikulti}) where work extraction is only zero at $p=1/2$ for $d=1$ and $d=2$. We  also note the recovery of the random access limit $\lim_{d\to\infty}P^{s}_u(d,1/2)=1$. Note that this first critical dimension  mirrors the critical dimension separating recurrence and transience in random walks due to Poly\'a \cite{polya_uber_1921}.  It is instructive to emphasise that the underlying behaviour behind this classification, the probability of escape to infinity, is the central reason for differences between dimensionalities. Intuitively as the dimensionality increases, the probability of escape also increases due to the increasing  multiplicity of possible paths. Similarly, as the dimension increases the walker simply has more physical space in which to diffuse that avoids its previous locations. Indeed a walker that \emph{never} resamples visited sites would trace a \emph{self-avoiding walk} which, for similar reasons, are superdiffusive on low dimensional topologies, but recover simple random walk diffusive behaviour in high dimensions \cite{madras_self-avoiding_1996}.
\par
Finally, we turn to the question of determining the optimal bias that maximizes work extraction from the information reservoir.  To understand this, we  consider the response of $\langle \dot{W}_{\text{net}}\rangle$ to small perturbations about $p=1/2$. Since the above integral is not analytic at $p=1/2$ this requires a  careful asymptotic analysis, performed in Appendix \ref{smIII}. We find a second critical  value of $d\in\mathbb{R}_+$, dependent on model parameters, must lie in $(4,\sim 4.037)$ which divides two distinct behaviours. For integer dimension, $d>4$, the optimal demon is unbiased, exploring the reservoir diffusively. However for lower integer dimensions ($d\leq 4$) a small bias is always optimal mirroring the dimension at which walkers become strongly transient at zero bias on such structures, detailed in Appendix \ref{smIV}. Note, however, that for the case of  $d=4$  our analysis predicts an exceptionally small, indeed  numerically undetectable, bias,  such that for all practical purposes we may state that all bias is detrimental to work extraction on hyper-cubic structures. 
\par
The complete behavior of the quantities that characterize the performance of work extraction as $d$ varies is illustrated in Fig.~(\ref{d_crit}) for an optimal demon. The first critical value of $d$ can be observed in the behaviour of $P_u^s(d,1/2)$ which departs the axis at $d=2^+$ leading to distinct behaviour in the integer dimensions $d=2$ and $d=3$. The second critical value of $d$ can be observed in the behaviour of $p_{\rm opt}$ which is non-zero only below $d\sim 4.037$ leading to distinct behaviour in the integer dimensions $d=4$ and $d=5$. And finally, optimal performance is captured by $\eta$ which increases markedly with increasing $d$, recovering $\eta=1$ in the limit $d\to\infty$ where random access is achieved for a purely diffusive walker, thus returning the behaviour of the BS model. 

\section{Discussion and Conclusion.}

 In this work we have investigated the thermodynamic consequences of modelling an information ratchet exploiting an information reservoir treated as a physical entity in its own right, possessing a defined spatial structure. Crucially, by treating the exploration of the reservoir as a thermal process we have identified a requirement for work input in order to efficiently acquire access to the low entropy resource. The required work input and the effectiveness of such a strategy has been found to depend heavily on the dimensionality of the information reservoir when modelled as a $d$-dimensional cubic lattice, characterised here by the existence of two critical dimensions. In particular, performance on one dimensional information reservoirs, the classically envisaged tape encoded with bits, is especially poor.
\par
As such we conclude that treating an information reservoir only through its statistical properties, i.e. its entropy, is not sufficient to characterize its potential for exploitation. The practical usefulness of an information reservoir as a resource depends crucially on its spatial structure. Only as the dimension becomes large do we approach a random access limit where the entropy is an appropriate characterization of its potential. We conjecture that this strong dependence of the net-extraction of work is the underlying reason why biological systems are not able to exploit polymers as information reservoirs which would have allowed them to recover some of the cost of synthesis.  
\par
 It is important to note the various idealisations made in this work. Firstly, our model has assumed that at each visit the demon equilibrates with the symbol on the currently bound site of the information reservoir. It may be interesting to relax this constraint in future work, however such an idealisation does not detract from the main conclusions of the paper as such a limit corresponds to maximal performance, where all of an information reservoir site's thermodynamic resource is consumed in one visit. Whilst resampling of sites with partially consumed resources, which would occur outside of timescale separation, will indeed contribute to work extraction to some extent, they too will be eventually consumed if new reservoir sites are not visited such that the delivery of new sites will remain the major performance bottleneck. 
\par
 Secondly, we have concerned ourselves with specific structures of hypothetical information reservoirs in this work, namely simple cubic lattices where the difference in structure is captured by the dimensionality. Generally, however, not all variation in performance between arbitrary structures would be characterised exclusively through their dimensionality. Despite this, we do expect the qualitative performance on any physically realisable crystalline structure to be heavily dependent on its dimensionality as the general properties of random walkers with respect to recurrence and transience, which are directly implicated in the ability to extract work, are robust with respect to individual lattice structures. That said, more exotic structures, such as graph structures that allow for arbitrary coordination numbers, those which restrict loops and so on, may be poorly characterised by their dimensionality and thus may have non-trivial dependencies in this regard.  It is unclear  whether such structures might be physically realisable. 
\par
Finally, throughout our analysis we assumed that the demon samples from visited and unvisited sites according to the average statistics of the random walk.  The model thus captures the mean behavior of the demon.  When observed over finite time-windows then the net-extraction of work will fluctuate even as $t\to\infty$. These fluctuations are not captured by our model, but they will disappear as the work extraction is averaged over sufficiently long time windows. 
\par
 More broadly, our results have implications for the thermodynamics of generalised information ratchets and models of information processing devices based upon them. This is because such ratchets only obtain maximal performance when operating upon information reservoirs without spatial structure. But in this limit only a bias of symbols in the reservoir's alphabet can be exploited.  However, it has been suggested \cite{merhav_relations_2017,merhav_sequence_2015}  that spatial correlations are exploitable by demons.  
We conjecture  that this is difficult. In order to detect reliably such correlations, directed motion is essential, leading to a cost of biasing the exploration of the information reservoir. In addition  demons that exploit spatial correlations  would  also have to bear the cost of ``computing'' a prediction of  future symbols based  on previously encountered ones \cite{boyd_identifying_2016,boyd_transient_2017}. This computation  comes at a finite cost and  leaves only very limited scope for  net-extraction of work. More generally, information processing utilising a specifically encoded set of instructions or upon specific data appears difficult to conceive of unless encoded upon a physical, spatially structured, substrate. Indeed such notions are most easily conceived of using one dimensional structures; precisely where performance is least optimal. We anticipate such a cost may, therefore, be prominent in broader microscopic models of computation and information processing more generally.

\begin{appendix}
\section{Treatment of the $1D$ tape}
\label{smI}
\subsection{Full transient solution}
As described in the main text, ${P}_u(1,p,n)$ is the probability that a random walker transitions to an unvisited site on the $n$th transition. This is simply related to the expected number of unique sites, $S(n)$, during a random walk of length $n$ and  the first passage probability $F(x,j|0,0)$ at site $x$ and time $j$, having started at the origin $0$ at time $0$.  To do so we have a fundamental relation between the expected number of unique sites and first passage probabilities
\begin{align}
S(n) = 1+\sum_{x\neq 0}   \sum_{j=1}^n F(x,j|0,0),
\label{Sn}
\end{align}
where we emphasise that $F(x,j|0,0)$ is the probability of \emph{arrival} at site $x$ (from another site) at time $j$ such that $F(0,0|0,0)=0$, and where the sum over $x$ (less the origin) is over all lattice sites, $\mathbb{Z}$. Alongside the occupation probability $P(x,j|0,0)$ to be at site $x$ at time $j$ having started at the origin $0$ at time $0$, we define the generating functions
\begin{align}
 H(x,z)&=\sum_{k=0}^{\infty} F(x,k|0,0)z^k\\
  G(x,z)&=\sum_{k=0}^{\infty} P(x,k|0,0)z^k.
 \end{align}
 A crucial relation we shall utilise is that between the above two generating functions \cite{redner_guide_2007}. This arises from the property that the probabilities of occupation and first passage $P(x,j|0,0)$ and $F(x,j|0,0)$ are related by
 \begin{align}
 P(x,j|0,0)&=\delta_{x,0}\delta_{j,0}+\sum_{k=1}^jF(x,k|0,0)P(0,j-k|0,0),
 \end{align}
 since all occupation events can be represented as first passage to the relevant site followed by net movement of $0$, assuming a time homogeneous process. 
  Recognising the convolution and the properties of generating functions/Laplace transforms we therefore have
\begin{align}
H(x,z)=\frac{G(x,z)-\delta_{x,0}}{G(0,z)}.
\label{rel0}
\end{align}
Considering $\Delta_k=S(k+1)-S(k)$, adapting \cite{montroll_random_1965}, with generating function $\Delta(z)$, we have, since $F(x,0|0,0)=0$,
\begin{align}
\Delta_k&= \sum_{x\neq 0}  F(x,k+1|0,0)\nonumber\\
\Delta(z)&=\sum_{k=0}^{\infty}\Delta_k z^k\nonumber\\
&=\left(\sum_{x}z^{-1}H(x,z)\right)-z^{-1}H(0,z).
\end{align}
Eq.~(\ref{rel0}) then gives 
\begin{align}
\Delta(z)&=\left(\sum_x\frac{G(x,z)}{zG(0,z)}\right)-\frac{1}{zG(0,z)}-\frac{G(0,z)-1}{zG(0,z)},
\end{align}
which along with normalisation and $\sum_{i=0}^{\infty} z^i = 1/(1-z),\ z<1$, allows us to write
\begin{align}
\Delta(z)&=\frac{1-G(0,z)(1-z)}{z(1-z)G(0,z)}.
\end{align}
Consequently, by considering 
\begin{align}
S(n)=\begin{cases}1+\sum_{i=0}^{n-1}\Delta_i & n>0\\
1&n=0,
\end{cases}
\end{align}
and recognising the more general $\sum_{i=k}^{\infty} z^i = z^k/(1-z),\ z<1$, we can write the generating function of $S(n)$ as
\begin{align}
S(z)&=\sum_{j=0}^{\infty}S(j)z^j\nonumber\\
&=S(0)+S(1)z+S(2)z^2+\ldots\nonumber\\
&=1+(1+\Delta_0)z+(1+\Delta_0+\Delta_1)z^2+\ldots\nonumber\\
&=\sum_{j=0}^\infty z^j+\Delta_0\sum_{j=1}^\infty z^j+\Delta_1\sum_{j=2}^\infty z^j+\ldots\nonumber\\
&=\frac{1}{1-z}+\frac{z\Delta_0}{1-z}+\frac{z^2\Delta_1}{1-z}+\ldots\nonumber\\
&=\frac{1}{1-z}+\frac{z\Delta(z)}{1-z}
\end{align}
allowing us finally to write
 \begin{align}
 S(z) = \frac{1}{G(0,z) (1-z)^2}.
 \end{align}
Now, for the discrete 1d random walk we have \cite[Eq.~(1.3.11)]{redner_guide_2007}
 \begin{align}
 G(x,z)=(1-4p(1-p)z^2)^{-1/2}\left(\frac{1-4p(1-p)z^2}{2z(1-p)}\right)^{|x|}
 \end{align}
   with $G(0,z)=(1-4p(1-p)z^2)^{-1/2}$. By substitution one can then evaluate 
\begin{align}
S(n) =\frac{1}{n!}\left.\frac{\partial^n}{ \partial z^n}S(z)\right|_{z=0}
\end{align}
which after repeated differentiation yields a geometric expression in terms of Pochhammer numbers which may be expressed
\begin{align}
 S(n) &=  \sum_{i=1}^{\ceil*{(n+1)/2}}  \frac{(2i-n-3)}{2\sqrt{\pi}}(i)_{-\frac{3}{2}} (4p(1-p))^{i-1},
\end{align}
where $(a)_k$ is the Pochhammer symbol
\begin{align}
(a)_k&=\frac{\Gamma[a+k]}{\Gamma[a]}
\end{align}
and $\Gamma[\cdot]$ the gamma function. 
We then compute the discrete derivative $dS(n+1)/dn=S(n+1)-S(n)$  finding 
 \begin{align}
&  \frac{dS(n{+}1)}{dn}\nonumber\\
&= \sum_{i=1}^{\ceil*{{(n+2)}/ 2 }}\frac{(2i-n-4)}{2\sqrt{\pi}}(i)_{-\frac{3}{2}} (4p(1-p))^{i-1}\nonumber\\
&\quad-  \sum_{i=1}^{\ceil*{{(n+1)}/ 2}}\frac{(2i-n-3)}{2\sqrt{\pi}}(i)_{-\frac{3}{2}} (4p(1-p))^{i-1}\nonumber\\
 &=\sum_{i=1}^{\ceil*{{(n+1)}\over 2} }-\frac{\Gamma[i-3/2](4p(1-p))^{i-1}}{2\sqrt{\pi}\Gamma[i]}\nonumber\\
 &\quad+
 \begin{cases}\frac{((2\ceil*{{n+2}\over 2} )-n-4)}{2\sqrt{\pi}}\frac{\Gamma[\ceil*{{(n+2)}/ 2}-3/2]}{\Gamma[\ceil*{{(n+2)}/ 2}]} & n\;\text{odd}\\
 \times (4p(1-p))^{\ceil*{{(n+2)}/ 2}-1}&\\
 0 & n\;\text{even}
 \end{cases}\nonumber\\
  &=\sum_{i=1}^{\ceil*{{(n)}\over 2}+1 }-\frac{\Gamma[i-3/2](4p(1-p))^{i-1}}{2\sqrt{\pi}\Gamma[i]},
  \end{align}
  such that
  \begin{align}
    \frac{dS(n)}{dn} &=\sum_{i=1}^{\ceil*{{(n+1)}\over 2} }-\frac{\Gamma[i-3/2](4p(1-p))^{i-1}}{2\sqrt{\pi}\Gamma[i]}.
  \end{align}
  We  now write $m=\ceil*{(n+1)/2}$ and consider the expression as the sum of two contributions
  \begin{align}
   \frac{dS(n)}{dn}&=\sum_{i=1}^{\infty}-\frac{\Gamma[i-3/2](4p(1-p))^{i-1}}{2\sqrt{\pi}\Gamma[i]}\nonumber\\
   &\quad-\sum_{i=m+1}^{\infty}-\frac{\Gamma[i-3/2](4p(1-p))^{i-1}}{2\sqrt{\pi}\Gamma[i]}\nonumber\\
   &=\sum_{i=0}^{\infty}-\frac{\Gamma[i-1/2](4p(1-p))^{i}}{2\sqrt{\pi}\Gamma[i+1]}\nonumber\\
   &\quad-\sum_{i=0}^{\infty}-\frac{\Gamma[i+m-1/2](4p(1-p))^{i+m}}{2\sqrt{\pi}\Gamma[i+m+1]}.
   \label{twosums}
  \end{align}
  We may now utilise the following identities
  \begin{align}
  \label{ident1}
  \sum_{i=0}^\infty\frac{\Gamma[i+n/2]}{\Gamma[i+1]}z^i&=\frac{\Gamma[n/2]}{(\sqrt{1-z})^n}\\
    \label{ident2}
    \sum_{i=0}^\infty\frac{(a)_i(b)_i z^i}{(c)_i i!}&={}_2F_1(a,b;c;z),
  \end{align}
  where $_2F_1$ is the hyper-geometric function. 
  \par
  By Eq.~(\ref{ident1}) the first term in Eq.~(\ref{twosums}) is identifiable as $\sqrt{1-4p(1-p)}=|1-2p|$ which equals the entire sum in the $n\to\infty$ limit. We may write the second as
  \begin{align}
 & \sum_{i=0}^{\infty}-\frac{\Gamma[i+m-1/2](4p(1-p))^{i+m}}{2\sqrt{\pi}\Gamma[i+m+1]}\nonumber\\
 &\quad=  \frac{(4p(1-p))^{m}}{2\sqrt{\pi}}\frac{\Gamma[m-1/2]}{\Gamma[m+1]}\nonumber\\
 &\quad\times\sum_{i=0}^{\infty}-\frac{\Gamma[i+1]\Gamma[m+1]\Gamma[i+m-1/2]}{\Gamma[i+1]\Gamma[m-1/2]\Gamma[i+m+1]}(4p(1-p))^{i}\nonumber\\
  &\quad= - \frac{(4p(1-p))^{m}}{2\sqrt{\pi}}\frac{\Gamma[m-1/2]}{\Gamma[m+1]}\nonumber\\
  &\quad\times\sum_{i=0}^{\infty}\frac{\Gamma[i+1]}{\Gamma[1]}\frac{\Gamma[i+m-1/2]}{\Gamma[m-1/2]}\frac{\Gamma[m+1]}{\Gamma[i+m+1]}\frac{(4p(1-p))^{i}}{\Gamma[i+1]}\nonumber\\
  &\quad=  -\frac{(4p(1-p))^{m}}{2\sqrt{\pi}}\frac{\Gamma[m-1/2]}{\Gamma[m+1]}\nonumber\\
  &\quad\qquad\times{}_2F_1(1,m-1/2;m+1;4p(1-p))
  \end{align}
  identifying the form in Eq.~(\ref{ident2}). Combining the two expressions and appreciating that the probability of encountering an unvisited site, in this 1D system, with bias $p$ at timestep $n$ is identical to $dS(n)/dn$ we finally have
  \begin{align}
& P_u(1,p,n) =|1{-}2p|\nonumber\\
&+\frac{(4p(1-p))^{m}\Gamma\left[m{-}\frac{1}{2}\right] {}_2F_1\left(1,m{-}\frac{1}{2};m{+}1;4p(1{-}p)\right)}{2\sqrt{\pi}\Gamma\left[m+1\right]}.
 \end{align}
 \par
  The above result is valid for a discrete time random walker. To convert to a Markov, and thus Poisson, continuous time process we write
  \begin{align}
  P_u(1,p,t)&=\sum_{n=0}^{\infty}e^{-\gamma t}\frac{(\gamma t)^n}{n!}P_u(1,p,n)
  \label{sumcont}
  \end{align}
 where $\gamma=r^++r^-$ is the escape rate of the continuous time random walker and $p=r^+/(r^++r^-)$. An appeal to the central limit theorem for large $n$ yields $P_u(1,p,t)\simeq P^d_u(1,p,n)|_{n=\gamma t}$.
 \subsection{Particular results and asymptotics}
\begin{figure}
  \caption{Asymptotic approximation of Eq.~(\ref{approx}) contrasted with explicit expression Eq.~(\ref{sumcont}) with $\gamma=1$.}
  \centering
    \includegraphics[width=0.47\textwidth]{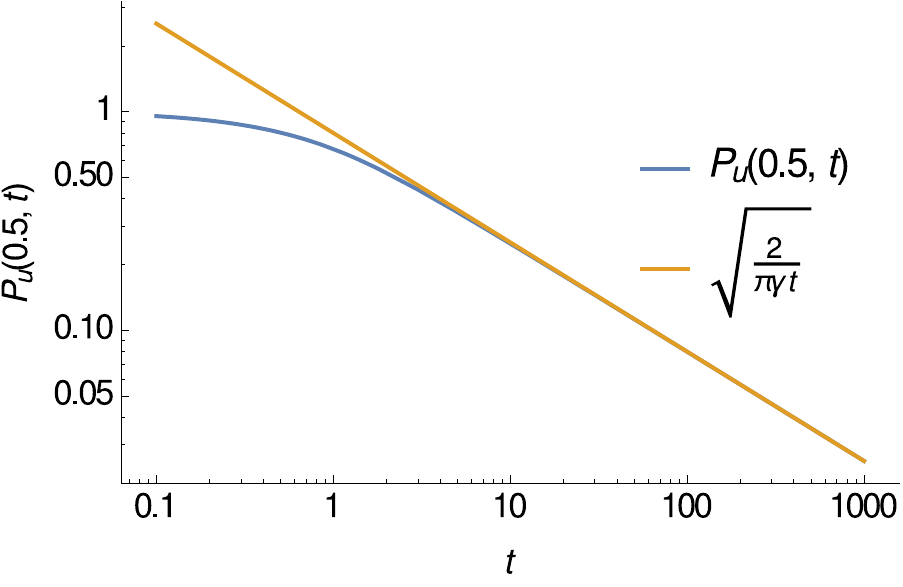}
    \label{1dea}
\end{figure}

\begin{figure}
  \caption{Behaviour of $P_u(1,p,t)-|1-2p|$ ranging from $p=0.55$ (top) to $p=0.9$ (bottom) in steps of $0.05$ in $p$ with $\gamma=1$.}
  \centering
    \includegraphics[width=0.47\textwidth]{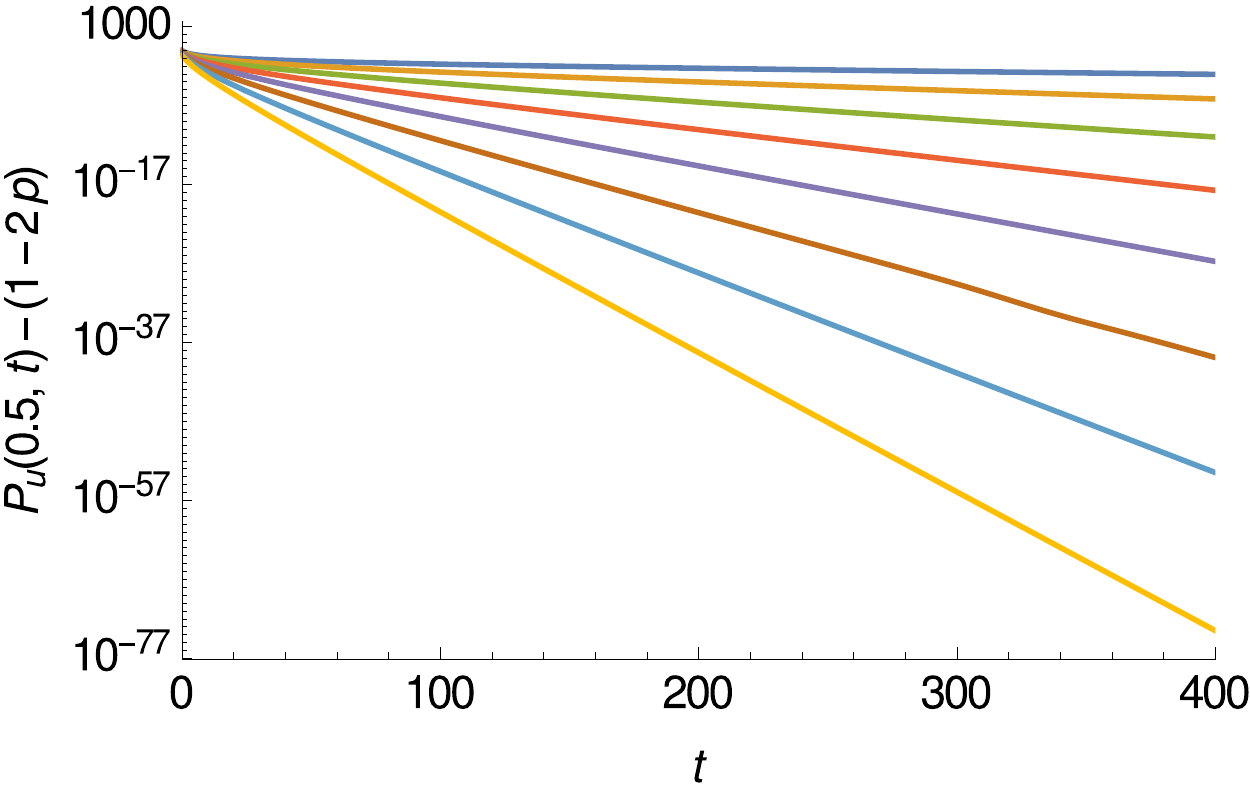}
    \label{exp_p}
\end{figure}

 A particular solution follows for $p=1/2$. Since for $p=1/2$ we have the form ${}_2F_1\left(1,m{-}\frac{1}{2};m{+}1;1\right)=2m$ we may simplify
 \begin{align}
 P^d_u(1,1/2,n)&=\frac{\Gamma[m-1/2]}{\sqrt{\pi}\Gamma[m]},
 \end{align}
 again with $m=\ceil*{(n+1)/2}$. 
 Furthermore, since we may represent the Gamma function with large argument $\Gamma[z]\simeq\sqrt{2\pi}z^{z-1/2}e^{-z}$ it follows that for large $n$ we may write
 \begin{align}
  P^d_u(1,1/2,n)&\simeq \left(\pi\ceil*{\frac{n+1}{2}}\right)^{-1/2}\nonumber\\
  &\simeq \sqrt{\frac{2}{n\pi}}.
  \end{align}
 When combined with the continuous time conversion formula, Eq.~(\ref{sumcont}), we find the asymptotically exact result for $p=1/2$,
 \begin{align}
   P_u(1,p,t)&\simeq \sqrt{\frac{2}{\pi\gamma t}},
   \label{approx}
   \end{align}
   which demonstrably exhibits power law decay with $t$. The accuracy of such an approximation against Eq.~(\ref{sumcont}) is shown in Fig.~(\ref{1dea}).
   The behaviour for $p\neq1/2$ is shown in Fig.~(\ref{exp_p}) demonstrating exponential decay with increasing speed as $p\to 0,1$.
\subsection{Optimal and critical parameters for the $1D$ reservoir}
\label{smIC}
As shown in the main text the work extracted from the heat reservoir has the form
\begin{align}
\gamma^{-1}\langle \dot{W}_{\rm ext}\rangle &=\frac{(1-\epsilon(1+e^{\frac{\Delta E}{T}}))\Delta E P_u(1,p,t)  }{(1+e^{\frac{\Delta E}{T}})}.
\end{align}
This is trivially maximised with respect to $\epsilon$ by setting $\epsilon=0$ leaving
\begin{align}
\gamma^{-1}\langle\dot{W}_{\rm ext}\rangle &=\frac{\Delta E  P_u(1,p,t)  }{(1+e^{\frac{\Delta E}{T}})}.
\label{www}
\end{align}
This maximises with respect to $\Delta E$ when
\begin{align}
\frac{e^{\frac{\Delta E}{T}}\Delta E}{(1+e^{\frac{\Delta E}{T}})T}=1,
\end{align}
through simple differentiation, which has solution corresponding to maximal work extraction, in terms of the Lambert W-function
\begin{align}
\Delta E&=T(1+\mathcal{W}(e^{-1}))
\end{align}
which upon substitution back into Eq.~(\ref{www}) is simply
\begin{align}
\gamma^{-1}\langle \dot{W}_{\rm ext}\rangle &=P_u(1,p,t)T\mathcal{W}(e^{-1})\simeq 0.27TP_u(1,p,t).
\end{align}
The critical bias, for this optimised demon, is then achieved by solving for $p$ when equating the work extracted from the reservoir and expended driving the tape along with $P_u(1,p,t)\to P^s_u(1,p)=|1-2p|$ such that
\begin{align}
|1-2p_c|T\mathcal{W}(e^{-1})&=T(2p_c-1)\ln\frac{p_c}{1-p_c}
\end{align}
which has solution ($p>1/2$)
\begin{align}
p_c&=(1+e\mathcal{W}(e^{-1}))^{-1}.
\end{align}
The optimal $p=p_{\rm opt}$ is achieved when the difference $\gamma^{-1}\langle \dot{W}_{\rm ext}\rangle-\gamma^{-1}\langle \dot{W}_{\rm tape}\rangle$ is maximised. Again, through simple differentiation of the relevant expressions, we obtain the relation 
\begin{align}
&p_{\rm opt}\nonumber\\
&=(1-p_{\rm opt})\exp\left[\mathcal{W}(e^{-1})+\frac{1}{2}\left(p_{\rm opt}^{-1}-(1-p_{\rm opt})^{-1}\right)^{-1}\right].
\end{align}
The ratio $\eta$ characterises the relative performance of a demon on a structured reservoir in contrast to one with random access and is defined as
\begin{align}
\eta&=\frac{\max_{p,\Delta E,\epsilon}\left[\langle \dot{W}_{\rm ext}\rangle-\langle \dot{W}_{\rm tape}\rangle\right]}{\max_{\Delta E,\epsilon}\left.\langle \dot{W}_{\rm ext}\rangle\right|_{P_u^s=1}}\nonumber\\
&=\left.\frac{\langle \dot{W}^{p=p_{\rm opt}}_{\rm ext}\rangle-\langle \dot{W}^{p=p_{\rm opt}}_{\rm tape}\rangle}{\langle \dot{W}^{P_u^s=1}_{\rm ext}\rangle}\right|_{\Delta E=T(1+\mathcal{W}(e^{-1})),\epsilon=0}\nonumber\\
&=\frac{T|1-2p_{\rm opt}|\mathcal{W}(e^{-1}) - T(2p_{\rm opt}-1)\ln\frac{p_{\rm opt}}{1-p_{\rm opt}}}{T\mathcal{W}(e^{-1})}\nonumber\\
&=(2p_{\rm opt}-1)\left(1+\mathcal{W}(e^{-1})^{-1}\ln\frac{p_{\rm opt}}{1-p_{\rm opt}}\right),
\end{align}
with the last line corresponding to the $p>1/2$ solution.
\section{Calculation of $P^s_u(d,p)$ for high dimensional reservoirs $d>1$}
\label{smII}
\subsection{$d$-dimensional, $\mathbb{Z}^d$, cubic lattices}
Following Montroll \cite{montroll_random_1956} we consider the discrete time random walk in dimension $d$ on $\mathbb{Z}^d$ (linear, square, cubic and  hyper-cubic lattices). We consider the points $\mathbf{x}=(x_1,x_2,\ldots,x_d)$ on a lattice with $N^d$ points in total such that $x_1\in\{1,2,\ldots,N\}$. We then write the probability of being at lattice point $\mathbf{x}$ at time $n$ to be $P(\mathbf{x};n)$. Our random walk, characterised by a uniform bias in all dimensions, is then given by the difference equation
\begin{align}
&P(x_1,x_2,\ldots,x_d,n+1)\nonumber\\
&\quad=\sum_{i=1}^d \frac{p}{d}P(x_1,\ldots,x_i-\epsilon_i,\ldots,x_d,n)\nonumber\\
&\qquad\qquad+\frac{(1-p)}{d}P(x_1,\ldots,x_i+\epsilon_i,\ldots,x_d,n),
\label{dif}
\end{align}
where $\epsilon_i\in\{-1,1\}$ controls the direction of bias along each of the dimension axes. 
This is then solved, subject to periodic boundary conditions $P(x_1+m_1N,x_2+m_2N,\ldots,x_d+m_dN,n)=P(x_1,x_2,\ldots,x_d,n)$ where $m_i\in\mathbb{N}$. These equations can be solved in a linear basis of terms
\begin{align}
b^n(\mathbf{l})N^{-\frac{d}{2}}\exp{\left[\frac{2\pi i}{N}(\mathbf{l}\cdot\mathbf{x})\right]}
\end{align}
where $\mathbf{l}=(l_1,l_2,\ldots,l_d)$ is a lattice point such that $l_i \in \{1,2,\ldots,N\}$. Directly substituting into Eq.~(\ref{dif}) gives
\begin{widetext}
\begin{align}
P(x_1,x_2,\ldots,x_d,n+1)&\sim b^{n+1}(\mathbf{l})N^{-\frac{d}{2}}\exp{\left[\frac{2\pi i}{N}(\mathbf{l}\cdot\mathbf{x})\right]}\nonumber\\
&=\sum_{i=1}^{d}d^{-1}pb^{n}(\mathbf{l})N^{-\frac{d}{2}}\exp{\left[\frac{2\pi i}{N}(l_1x_1,\ldots,l_i(x_i-\epsilon_i),\ldots,l_dx_d)\right]}\nonumber\\
&\quad\quad+d^{-1}(1-p)b^{n}(\mathbf{l})N^{-\frac{d}{2}}\exp{\left[\frac{2\pi i}{N}(l_1x_1,\ldots,l_i(x_i+\epsilon_i),\ldots,l_dx_d)\right]}
\end{align}
\end{widetext}
giving
\begin{align}
b(\mathbf{l})&=\sum_{i=1}^dd^{-1}p\exp{\left[-\frac{2\pi i}{N}l_i\epsilon_i\right]}+(1-p)\exp{\left[\frac{2\pi i}{N}l_i\epsilon_i\right]}\nonumber\\
&=\sum_{i=1}^dd^{-1}\cos{\left(\frac{2\pi l_i}{N}\right)}+i\epsilon_id^{-1}(1-2p)\sin{\left(\frac{2\pi l_i}{N}\right)}
\end{align}
such that
\begin{align}
P(\mathbf{x},n)&=\sum_{\mathbf{l}}a(\mathbf{l})b^{n}(\mathbf{l})N^{-\frac{d}{2}}\exp{\left[\frac{2\pi i}{N}(\mathbf{l}\cdot\mathbf{x})\right]}
\end{align}
where we note that the sum over $\mathbf{l}$ is shorthand for
\begin{align}
\sum_{\mathbf{l}}\equiv\sum_{l_1=1}^N\ldots \sum_{l_d=1}^N.
\label{defsum}
\end{align}
$a(\mathbf{l})$ is then determined by considering an initial condition $P(\mathbf{x},0)$. Writing
\begin{align}
P(\mathbf{x},0)&=\sum_{\mathbf{y}}P(\mathbf{y},0)\delta(\mathbf{x}-\mathbf{y})\nonumber\\
&=N^{-d}\sum_{\mathbf{y}}P(\mathbf{y},0)\sum_{\mathbf{l}}\exp{\left[\frac{2\pi i}{N}\mathbf{l}\cdot(\mathbf{x}-\mathbf{y})\right]}\nonumber\\
&=N^{-d}\sum_{\mathbf{l}}\left\{\sum_{\mathbf{y}}P(\mathbf{y},0)\exp{\left[-\frac{2\pi i}{N}\mathbf{l}\cdot\mathbf{y}\right]}\right\}\nonumber\\
&\qquad\qquad\times\exp{\left[\frac{2\pi i}{N}\mathbf{l}\cdot\mathbf{x}\right]}\nonumber\\
&=\sum_{\mathbf{l}}a(\mathbf{l})N^{-\frac{d}{2}}\exp{\left[\frac{2\pi i}{N}(\mathbf{l}\cdot\mathbf{x})\right]}
\end{align}
shows
\begin{align}
a(\mathbf{l})&=N^{-\frac{d}{2}}\sum_{\mathbf{y}}P(\mathbf{y},0)\exp{\left[-\frac{2\pi i}{N}\mathbf{l}\cdot\mathbf{y}\right]},
\end{align}
where the sum over the lattice points $\mathbf{y}$ follows Eq.~(\ref{defsum}). This then allows us to write the probability of being at a lattice site $\mathbf{x}$ having started at some other point $\mathbf{x}'$, at time $0$ as
\begin{widetext}
\begin{align}
P(\mathbf{x},n|\mathbf{x}',0)&=N^{-d}\sum_{\mathbf{l}}\left[\sum_{i=1}^dd^{-1}\cos{\left(\frac{2\pi l_i}{N}\right)}+i\epsilon_id^{-1}(1-2p)\sin{\left(\frac{2\pi l_i}{N}\right)}\right]^n\exp{\left[\frac{2\pi i}{N}\mathbf{l}\cdot(\mathbf{x}-\mathbf{x}')\right]}.
\end{align}
Letting $N\to \infty$ allows expression as a $d$ dimensional integral
\begin{align}
P(\mathbf{x},n|\mathbf{x}',0)&=(2\pi)^{-d}\underbrace{\int_{0}^{2\pi}\ldots\int_{0}^{2\pi}}_{\text{d times}}\left[d^{-1}\sum_{i=1}^d\cos{\left(\theta_i\right)}+i\epsilon_i(1-2p)\sin{\left(\theta_i\right)}\right]^n\exp{\left[i\mathbf{\theta}\cdot (\mathbf{x}-\mathbf{x}')\right]}\ d\theta_1,\ldots,d\theta_d
\end{align}
where $\mathbf{\theta}=(\theta_1,\theta_2,\ldots,\theta_d)=(2\pi l_1/N,2\pi l_2/N,\ldots,2\pi l_d/N)$. Relabelling $\mathbf{x}-\mathbf{x}'\to\mathbf{x}$ then allows consideration of having started at the origin, $\mathbf{o}$, viz.
\begin{align}
P(\mathbf{x},n|\mathbf{o},0)&=(2\pi)^{-d}\underbrace{\int_{0}^{2\pi}\ldots\int_{0}^{2\pi}}_{\text{d times}}\left[d^{-1}\sum_{i=1}^d\cos{\left(\theta_i\right)}+i\epsilon_i(1-2p)\sin{\left(\theta_i\right)}\right]^n\exp{\left[i\mathbf{\theta}\cdot \mathbf{x}\right]}\ d\theta_1,\ldots,d\theta_d.
\end{align}
\par
Using the above and the property $\sum_{n=0}^{\infty}a^n=(1-a)^{-1},\ 0<a<1$, we can express the generating function
\begin{align}
G(\mathbf{x},z)&=\sum_{n=0}^{\infty}P(\mathbf{x},n|\mathbf{o},0)z^{n}
\end{align}
as
\begin{align}
G(\mathbf{x},z)&=\frac{1}{(2\pi)^d}\underbrace{\int_{0}^{2\pi}\ldots\int_{0}^{2\pi}}_{\text{d times}}\frac{\exp{\left[i\mathbf{\theta}\cdot\mathbf{x}\right]}}{1-zd^{-1}\sum_{i=1}^d\cos{\left(\theta_i\right)}+i\epsilon_i(1-2p)\sin{\left(\theta_i\right)}}d\theta_1,\ldots,d\theta_d
\end{align}
or, since we can write $(1-y)^{-1}=\int_0^\infty\exp[x(y-1)]dx$, $0<y<1$,
\begin{align}
G(\mathbf{x},z)&=\frac{1}{(2\pi)^d}\int_0^\infty e^{-x}\underbrace{\int_0^{2\pi}\ldots \int_0^{2\pi}}_{\text{d times}}e^{i\mathbf{x}\cdot \theta}e^{xzd^{-1}\left(\cos(\theta_i)+i\epsilon_i(1-2p)\sin(\theta_i)\right)}d\theta_1\ldots d\theta_d\ dx\nonumber\\
&=\int_0^\infty e^{-x}\prod_{i=1}^{d}\left[\frac{1}{2\pi}\int_0^{2\pi}e^{ix_i\theta_i}e^{xzd^{-1}\left(\cos(\theta_i)+i\epsilon_i(1-2p)\sin(\theta_i)\right)}d\theta_i\right]dx.
\label{genbessel}
\end{align}
\end{widetext}
The generating function with $\mathbf{x}=\mathbf{o}$, $G(\mathbf{o},z)$, then concerns the probability of being at the origin at time step $n$, having started at the origin at time $0$, $P(\mathbf{o},n|\mathbf{o},0)$. One can then consider the probability of returning to the origin for the \emph{first} time at time $n$, $P_r^f(d,p,n)=F(\mathbf{o},n|\mathbf{o},0)$, with corresponding generating function 
\begin{align}
H(\mathbf{o},z)=\sum_{n=0}^{\infty}P_r^f(d,p,n)z^{n}.
\end{align}
 Crucially, from Eq.~(\ref{rel0}), these generating functions are related as 
 \begin{align}
 G(\mathbf{o},z)-1=G(\mathbf{o},z)H(\mathbf{o},z).
 \end{align}
  By then considering the probability of \emph{ever} returning to the origin being 
  \begin{align}
  \sum_{n=0}^\infty P_r^f(d,p,n)=H(\mathbf{o},1),
  \end{align}
   one finds the probability of ever returning to the origin, the probability of recurrence, $P_r(d,p)$, to be given by 
   \begin{align}
   P_r(d,p)=1-(G(\mathbf{o},1))^{-1}.
   \label{reffinal}
\end{align}
 Here we appeal to the fact that $1-P_r(d,p)=\lim_{n\to\infty} dS(n)/dn$, where $S(n)$ is the expected number of sites visited at least once at time $n$,  stemming from the more general property $1-P_r(d,p,n-1)=dS(n)/dn$ originally due to Dvoretzky and Erd\"{o}s \cite{a._dvoretzky_problems_1951,vineyard_number_1963}. This can be seen by considering the occupation of site $\mathbf{x}$ at time $n$ as the sum of $n$ i.i.d. random vectors $\mathbf{v}_i$, $i\in\{1,\ldots,n\}$ such that $\mathbf{y}_m\equiv\{\mathbf{x}=\mathbf{y},n=m\}=\sum_{i=1}^m\mathbf{v}_i$. Using this description we may write
 \begin{align}
 \frac{dS(n)}{dn}&=P\left\{\sum_{i=1}^n\mathbf{v}_i\neq \sum_{i=1}^j\mathbf{v}_i,\ \forall j\in\{1,\ldots,n{-}1\}\right\}\nonumber\\
 &=P\left\{\sum_{i=j+1}^n\mathbf{v}_i\neq 0,\ \forall j\in\{1,\ldots,n{-}1\}\right\}\nonumber\\
  &=P\left\{\sum_{i=1}^{n-j}\mathbf{v}_i\neq 0,\ \forall j\in\{1,\ldots,n{-}1\}\right\}\nonumber\\
    &=P\left\{\sum_{i=1}^{j}\mathbf{v}_i\neq 0,\ \forall j\in\{1,\ldots,n{-}1\}\right\}\nonumber\\
        &=P\left\{\mathbf{x}_j\neq 0,\ \forall j\in\{1,\ldots,n{-}1\}\right\},
 \end{align}
 which is the probability of not returning to the origin between times $0$ and $n-1$, i.e. $1-P_r(d,p,n-1)$, and where we have assumed time homogeneity in going from line $2$ to line $3$.
\par 
Now, by definition $dS(n)/dn$ is also the mean field probability of discovering a new site, $P_u(d,p,n)$, such that $\lim_{n\to\infty}dS(n)/dn=P_u(d,p)$.   
Consequently, through Eq.~(\ref{reffinal}), we finally have $P_u^s(d,p)=(G(\mathbf{o},1))^{-1}$, where we now introduce the explicit quantity $u(d,p)=G(\mathbf{o},1)$. Despite being derived in discrete time, in the $t\to\infty$ limit the distinction becomes irrelevant  and thus, consequently, we have the expression for $P_u^s(d,p)=u^{-1}(d,p)$, for our continuous time walker,
\begin{align}
&(P_u^s(d,p))^{-1}=u(d,p)\nonumber\\
&=\underbrace{\int_{0}^{2\pi}\ldots\int_{0}^{2\pi}}_{\text{d times}}\frac{(2\pi)^{-d}d\theta_1\ldots d\theta_d}{1-d^{-1}\sum_{i=1}^d\cos{\left(\theta_i\right)}+i\epsilon_i(1-2p)\sin{\left(\theta_i\right)}}\nonumber\\
&=\int_0^\infty e^{-x}\prod_{i=1}^d\left[\frac{1}{2\pi}\int_0^{2\pi}e^{xd^{-1}\left(\cos(\theta_i)+i\epsilon_i(1-2p)\sin(\theta_i)\right)}d\theta_i\right]dx.
\label{int1}
\end{align}
Since $\epsilon_i(1-2p)$ must lie in $[-1,1]$ we may define $\cos(i\delta)=(1-(1-2p)^2)^{-1/2}$, $\sin(i\delta)=i\epsilon_i(1-2p)(1-(1-2p)^2)^{-1/2}$. Since $(\cos(b))^{-1}\cos(a+b)=\cos(a)+\sin(a)\tan(b)$ we have, letting $a\to\theta_i, b\to i\delta$, $\cos{\left(\theta_i\right)}+i\epsilon_i(1-2p)\sin{\left(\theta_i\right)}=2(p(1-p))^{1/2}\cos(\theta_i+i\epsilon_i\delta)$. Considering then that the rectangular contour integral $z=x+iy= i\epsilon_i\delta \to 2\pi+i\epsilon_i\delta\to 2\pi\to 0\to i\epsilon_i\delta$ contains no poles, $(P_u^s(d,p))^{-1}$, corresponding to the first edge must be independent of $\delta$ (since $\cos(\pm i\delta)=\cos(2\pi\pm i\delta)$) allowing us to set $\delta =0$ and thus recognise
\begin{align}
(P_u^s(d,p))^{-1}&=u(d,p)\nonumber\\
&=\int_0^\infty e^{-x}\left[I_0\left(\frac{2x}{d}\sqrt{p(1-p)}\right)\right]^d dx
\label{int2_app}
\end{align}
where $I_0$ is the zero-th modified Bessel function of the first kind, which follows from the integral representation $I_0(x)=\pi^{-1}\int_0^\pi e^{x\cos(\theta)}d\theta$. We note that for all $d$, if $p=0$ or $p=1$ then the integral reduces to $\int_0^{\infty} e^{-x}dx=1$ indicating an escape probability, and thus $P_u^s(d,p)$, of $1$ at total bias (complete irreversibility).
\par
The above integral has known solutions for $d=1$ and $d=2$
\begin{align}
(P_u^s(1,p))^{-1}&=\frac{1}{|1-2p|}\nonumber\\
(P_u^s(2,p))^{-1}&=\frac{2K[4p(1-p)]}{\pi}
\end{align}
where $K$ is the complete elliptic integral of the first kind and where the 1D result agrees with our result in the main text.
\subsection{Expressions for alternative lattice structures}
\label{AppB2}
For completeness we derive expressions for $P_u^s(d,p)$ for alternative lattices. We can once again follow Montroll \cite{montroll_random_1956} in such cases where the preliminary theory is essentially identical to the simple cubic case and recognise the more general form for the generating function
\begin{align}
G(\mathbf{x},z)&=\frac{1}{(2\pi)^d}\underbrace{\int_{0}^{2\pi}\ldots\int_{0}^{2\pi}}_{\text{d times}}\frac{\exp[i\theta\cdot\mathbf{x}]}{1-z\lambda(\theta)}d\theta_1\ldots d\theta_d
\end{align}
and thus
\begin{align}
(P_u^s)^{-1}&=\frac{1}{(2\pi)^d}\underbrace{\int_{0}^{2\pi}\ldots\int_{0}^{2\pi}}_{\text{d times}}\frac{1}{1-\lambda(\theta)}d\theta_1\ldots d\theta_d
\label{dfv}
\end{align}
where $\lambda(\theta)$ is a structure function given as
\begin{align}
\lambda(\theta)&=\sum_{\mathbf{x}}P(\mathbf{x},n+1|\mathbf{x}',n)\exp[i(\mathbf{x}-\mathbf{x}')\cdot\theta]
\end{align}
and where $\mathbf{x}'$ is arbitrary and can be chosen as the origin such that $P(\mathbf{x},n+1|\mathbf{x}',n)=P(\mathbf{x},1|\mathbf{o},0)$.
\par
Different lattices then have different structure functions. For simple cubic (sc) lattices they are defined by walkers that must move one step in only one dimension at each time step resulting in the form in the previous section.
\subsubsection{$d$-dimensional body-centred and face-centred cubic lattices}
$d$-dimensional bcc lattices are defined as two perfectly inter-woven $d$-dimensional simple cubic lattices such that a walker transitioning a fixed distance between nearest neighbours would do so along $2^{d-1}$ un-signed directions, or axes, (giving coordination number $2^d$). These $2^{d-1}$ different jump axes correspond to a walker which must move forwards or backwards one step along every dimension in every time step.
\par
$d$-dimensional fcc lattices can be defined as a simple hypercubic lattice with additional lattice points at the centre of every possible square face on that hypercube. This gives $d(d-1)$ axes for a walker to transition thus giving a coordination number of $2d(d-1)$. These $d(d-1)$ different jump axes correspond to a walker which must move forwards or backwards one step along exactly two dimensions in every time step. I.e. there are $_dC_2=(n/2)(n-1)$ planes a walker can transition in, each with four nearest neighbours corresponding to the coordination number of $2d(d-1)$.
\par
According to these definitions there is no distinction between sc, bcc and fcc structures for $d<3$ except for the nearest neighbour distance which plays no role here.
\par
We assume that when bias is introduced to the walker, it is introduced such that a) it transitions along one of the $2^{d-1}$ or $d(d-1)$ axes with equal probability and b) it transitions in one direction along such an axis with probability $p$ and the other with probability $(1-p)$.
\par
\begin{figure}
  \caption{Illustration of $P_u^s$ for the triangular and square lattices for any choice of bias direction other than $\epsilon_1=\epsilon_2=-\epsilon_{12}$ in the triangular lattice.}
  \centering
    \includegraphics[width=0.47\textwidth]{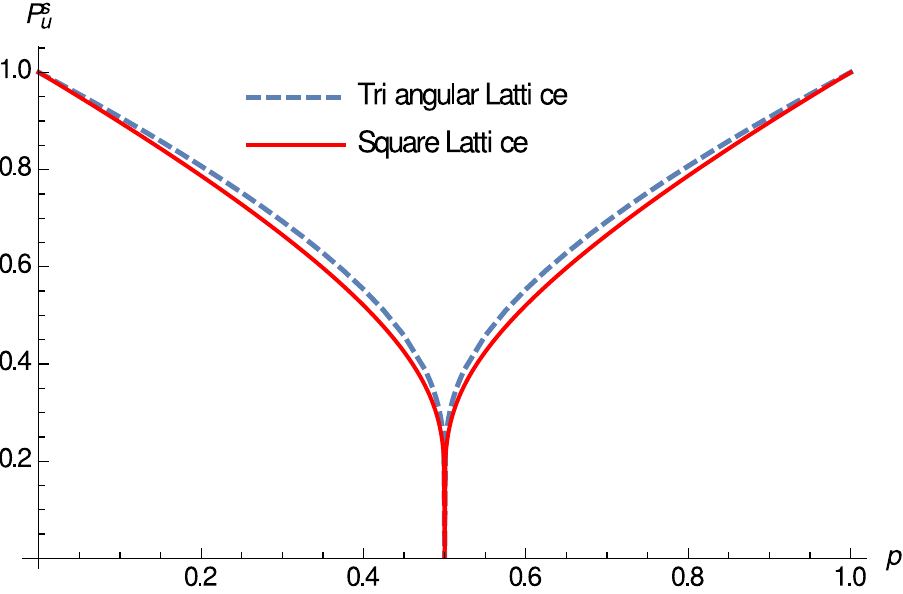}
    \label{square}
\end{figure}

Consequently the structure function for the $d$-dimensional bcc lattice is given by
\begin{widetext}
\begin{align}
\lambda_{\rm bcc}(d,\theta)&=2^{-(d+1)}\sum_{a_1\in\{-1,1\}}\ldots \sum_{a_d\in\{-1,1\}}\left[pe^{i\epsilon_{a_1\ldots a_d}(a_1\theta_1+a_2\theta_2+\ldots a_d\theta_d)}+(1-p)e^{-i\epsilon_{a_1\ldots a_d}(a_1\theta_1+a_2\theta_2+\ldots a_d\theta_d)}\right]
\end{align}
where $\epsilon_{a_1\ldots a_d}\in\{-1,1\}$ specifies whether bias is directed forwards or backwards along each jump axis. We note the double counting of each axis, and thus prefactor $2^{-(d+1)}$, since, for instance with $d=3$, the axis corresponding to $\epsilon_{1,-1,1}$ is the same as $\epsilon_{-1,1,-1}$ such that we also require $\epsilon_{a_1\ldots a_d}=-\epsilon_{a^-_1\ldots a^-_d}$ where $a_i^-=-a_i$.
\par
Similarly, the structure function for the $d$-dimensional fcc lattice is given by
\begin{align}
\lambda_{\rm bcc}(d,\theta)&=2(d(d-1))^{-1}\sum_{i=1}^d\sum_{j=i+1}^d \sum_{k\in\{-1,1\}}\left[pe^{i\epsilon_{{ijk}}(\theta_i+k\theta_j)}+(1-p)e^{-i\epsilon_{ijk}(\theta_i+k\theta_j)}\right].
\end{align}
where $\epsilon_{ijk}\in\{-1,1\}$ specifies the direction of bias along the axis $\theta_i+k\theta_j$ and where the double counting has been explicitly avoided.
\par
Importantly, unlike for the simple cubic structure, because the jump axes are not orthogonal, the resulting integrals retain dependence on the specific choices of $\epsilon_{a_1\ldots a_d}$ and $\epsilon_{ijk}$. Indeed for the same reason it does not follow that $P^s_u(d,1)=1$ under total bias for all choices of $\epsilon_{a_1\ldots a_d}$ and $\epsilon_{ijk}$.
\par
Choosing $d=3$ for illustrative purposes, multiple applications of the sum and difference trigonometric formulae reveals that the structure function for the bcc lattice is given by
\begin{align}
\lambda_{\rm bcc}(3,\theta)&=\cos(\theta_1)\cos(\theta_2)\cos(\theta_3)+\frac{i(2p-1)}{4}\left[\epsilon_{1,1,1}\sin(\theta_1+\theta_2+\theta_3)+\epsilon_{-1,1,1}\sin(-\theta_1+\theta_2+\theta_3)\right.\nonumber\\
&\qquad\left.+\epsilon_{1,-1,1}\sin(\theta_1-\theta_2+\theta_3)+\epsilon_{1,1,-1}\sin(\theta_1+\theta_2-\theta_3)\right].
\end{align}
Utilising the particular choice $\epsilon_{-1,1,1}=\epsilon_{1,-1,1}=\epsilon_{1,1,-1}=1$, $\epsilon_{1,1,1}=-1$ allows the simpler form
\begin{align}
\lambda_{\rm bcc}(3,\theta)&=\cos(\theta_1)\cos(\theta_2)\cos(\theta_3)-i(1-2p)\sin(\theta_1)\sin(\theta_2)\sin(\theta_3).
\end{align}
The structure function for the $d=3$ fcc lattice is given by
\begin{align}
\lambda_{\rm fcc}(3,\theta)&=\frac{1}{3}\left[\cos(\theta_1)\cos(\theta_2)+\cos(\theta_1)\cos(\theta_3)+\cos(\theta_2)\cos(\theta_3)\right]\nonumber\\
&+\frac{i(2p-1)}{6}\left[\epsilon_{1,2,1}\sin(\theta_1+\theta_2)+\epsilon_{1,2,-1}\sin(\theta_1-\theta_2)+\epsilon_{1,3,1}\sin(\theta_1+\theta_3)\right.\nonumber\\
&\qquad\left.+\epsilon_{1,3,-1}\sin(\theta_1-\theta_3)+\epsilon_{2,3,1}\sin(\theta_2+\theta_3)+\epsilon_{2,3,-1}\sin(\theta_2-\theta_3)\right].
\end{align}
Choosing $\epsilon_{1,2,1}=\epsilon_{1,2,-1}=\epsilon_{1,3,1}=\epsilon_{1,3,-1}=\epsilon_{2,3,1}=\epsilon_{2,3,-1}=1$ yields the simpler form
\begin{align}
\lambda_{\rm fcc}(3,\theta)&=\frac{1}{3}\left[\cos(\theta_1)\cos(\theta_2)+\cos(\theta_1)\cos(\theta_3)+\cos(\theta_2)\cos(\theta_3)\right]\nonumber\\
&+\frac{i(2p-1)}{3}\left[\cos(\theta_3)\sin(\theta_1)+\cos(\theta_2)\sin(\theta_1)+\cos(\theta_2)\sin(\theta_3)\right].
\end{align}
It should be emphasised that upon the choice $p=1/2$ the resultant integrals reduce to the known Watson triple integrals with well known relation to the sc, bcc and fcc structures. These have known results and so for the unbiased $p=1/2$ case we may establish that in contrast to the simple cubic result $P_u^{s,sc}(3,1/2)\simeq 0.65946$, we have $P_u^{s,bcc}(3,1/2)\simeq 0.71777$ and $P_u^{s,fcc}(3,1/2)\simeq 0.74368$ indicating that for $d=3$ and $p=1/2$ fcc structures outperform bcc structures which in turn outperform sc structures. It is unknown to us, and beyond scope, as to how to reduce the $d$ dimensional integrals as with the simple cubic lattices and thus we consider only simple cubic lattices in all further analysis.

\subsubsection{Triangular lattice}

Using the form in Eq.~(\ref{dfv}) it is straightforward to calculate the expression for $P_u^s$ for the two dimensional triangular lattice by mapping it to a a square lattice with additional diagonal connections along only one of the diagonals. This gives an expression
\begin{align}
(P_u^s)^{-1}&=\frac{1}{(2\pi)^2}\int_{0}^{2\pi}\int_{0}^{2\pi}\frac{d\theta_1d\theta_2}{1-\frac{1}{3}\left[\cos(\theta_1)+\cos(\theta_2)+\cos(\theta_1+\theta_2)+i(1-2p)(\epsilon_1\sin(\theta_1)+\epsilon_2\sin(\theta_2)+\epsilon_{3}\sin(\theta_1+\theta_2))\right]}
\end{align}
\end{widetext}
with $\epsilon_{i}\in\{-1,1\}$ corresponding to the three axes that make the triangular lattice. This integral has no obvious simplification, but is numerically tractable. For the particular case $\epsilon_1=\epsilon_2=-\epsilon_{3}$ the bias is symmetrically pointing in/out of any given lattice point (alternating biased and anti-biased directions towards each of the six nearest neighbours around the lattice point) such that the system becomes frustrated and the walker becomes recurrent for all values of $p$ such that $P_u^s=0$ for any bias. For all other choices the behaviour is independent of the values of $\epsilon_i$ and exhibits the behaviour shown in Fig.~(\ref{square}) which is qualitatively very similar to the square lattice, in particular demonstrating recurrence (and thus zero long term work extraction) at $p=1/2$. 

\section{Asymptotics and identification of critical dimensions for hyper cubic lattices}
\label{smIII}
Proceeding with the simple cubic lattice structure and the expression in Eq.~(\ref{int2_app}), we recognise that 
the modified Bessel functions of the first kind, $I_0(x)$, can be asymptotically approximated for large $x$ to be $I_0(x)\sim e^{x}/\sqrt{2\pi x}$. Consequently we may test convergence of $u(d,p)$ with a simple $p$-test \footnote{We recall the $p$-test identifies convergence in integrals of the form $\int_c^\infty f(x)dx$, if $0\leq f(x)\leq x^{-p}\; \forall x\in[c,\infty)$ when $c>0$ if $p>1$ resulting from the convergence of $\int_c^\infty x^{-p} dx$ for such values}. Noting that we can write the integral involving Bessel functions as
\begin{align}
&u(d,p)\nonumber\\
&=\frac{d}{2\sqrt{p(1-p)}}\int_0^{\infty}e^{-xd\left(\frac{1}{2\sqrt{p(1-p)}}-1\right)}\left[e^{-x}I_0(x)\right]^d dx
\label{eq1}
\end{align}
 we have
\begin{align}
&\lim_{c\to \infty}\frac{d}{2\sqrt{p(1-p)}}\int_c^\infty e^{-xd\left(\frac{1}{2\sqrt{p(1-p)}}-1\right)}\left[e^{-x}I_0(x)\right]^d dz\nonumber\\
&=\frac{d}{2\sqrt{p(1-p)}}\int_c^{\infty}\frac{e^{-xd\left(\frac{1}{2\sqrt{p(1-p)}}-1\right)}}{(2\pi x)^{\frac{d}{2}}}dx
\end{align}
which demonstrably diverges for $d=1$ and $d=2$ when $p=1/2$ with convergence otherwise indicating the first critical dimension, $d=3$, where work extraction can occur at zero bias.
\par
Looking towards the critical behaviour in whether zero bias is locally optimal, we consider the local behaviour in $u(d,p)$ and thus the work extracted in the region $p=1/2$. We might naively attempt to approximate by a truncated Taylor series where we would note that $du(d,p)/dp\big|_{p=1/2}=0$ for $d\geq 5$ and undefined elsewhere since the limit $\lim_{p\to 1/2}du(d,p)/dp$ diverges and that the second derivative at this point is similarly given by
\begin{align}
\frac{d^2u(d,p)}{dp^2}\Big|_{p=1/2}=-4\int_0^\infty ze^{-z}\left[I_0\left(\frac{z}{d}\right)\right]^{d-1}I_1\left(\frac{z}{d}\right)dz
\label{sd}
\end{align}
which by similar logic to the above can be shown to converge for $d\geq 5$ also. Attempting to expand the work extracted around this point gives
\begin{align}
&\gamma^{-1}\langle\dot{W}\rangle=\gamma^{-1}\langle\dot{W}\rangle_{p=1/2}\nonumber\\
&-\frac{1}{2}\frac{(1-\epsilon(1+e^{\frac{\Delta E}{T}}))\Delta E}{(1+e^{\frac{\Delta E}{T}})(u(d,1/2))^2}\frac{d^2u(d,p)}{dp^2}\Big|_{p=1/2}\left(p-{1}/{2}\right)^2\nonumber\\
&\qquad+\mathcal{O}\left(\left(p-{1}/{2}\right)^3\right).
\label{d5work}
\end{align}
The speed of growth in $p$ maximises for $\epsilon=0$, $\Delta E=T(1+\mathcal{W}(e^{-1}))$, where $\mathcal{W}$ is the Lambert W-function, such that
\begin{align}
\gamma^{-1}\langle\dot{W}\rangle&=\gamma^{-1}\langle\dot{W}\rangle_{p=1/2}\nonumber\\
&-\frac{\mathcal{W}(e^{-1})T}{2(u(d,1/2))^2}\frac{d^2u(d,p)}{dp^2}\Big|_{p=1/2}\left(p-{1}/{2}\right)^2\nonumber\\
&\qquad+\mathcal{O}\left(\left(p-{1}/{2}\right)^3\right).
\end{align}
For $d=5$ the quadratic term $\simeq 0.324 T$, $d=6$ it is $\simeq 0.177 T$ and decreases monotonically as $d$ increases. This \emph{may} be sufficient for $d\geq 5$, but fails for $d<5$ and moreover requires confirmation for $d\geq 5$ as faster terms may contribute from a proper asymptotic analysis.
\par
To achieve this we note that Eq.~(\ref{eq1}) is in the form $\int_0^\infty h(\lambda x)f(x) dx$ where $h(t)=e^{-t}$ and $\lambda = d((4p(1-p))^{-1/2}-1)$ reducing to a Laplace transform, i.e.
\begin{align}
u(d,p)&=(\lambda+d)\int_0^\infty e^{-\lambda x}\left[e^{-x}I_0(x)\right]^d dx
\label{ltform}
\end{align} 
 for which standard asymptotic integral techniques can be used where the limit $p\searrow 1/2$ corresponds to $\lambda\to 0^+$ \cite{norman_bleistein_asymptotic_1986,fikioris_integral_2006,wong_asymptotic_2001,wong_generalized_1984}. To proceed we require the following asymptotic forms
\begin{align}
f_d(t)&=\left[e^{-t}I_0(t)\right]^d\sim \sum_{s=0}^\infty c_s t^{-k_s}=(2\pi t)^{-\frac{d}{2}}\quad t\to\infty\nonumber\\
h(t)&=e^{-t}\sim \sum_{i=0}^{\infty}p_it^{a_i}=\sum_{i=0}^{\infty}\frac{(-1)^i}{i!}t^i\quad t\to 0^+
\label{asym}
\end{align}
and follow \cite{norman_bleistein_asymptotic_1986} in constructing the $h$-transform (here the Laplace transform) as the integral across the real line
 \begin{align}
 &\int_0^\infty h(\lambda t)f(t)dt\nonumber\\
 &=\frac{1}{2\pi i}\int_{r-i\infty}^{r+i\infty}M[h(\lambda t);1-z]M[f_d(t);z]dz\nonumber\\
 &=\frac{1}{2\pi i}\int_{r-i\infty}^{r+i\infty}\lambda^{z-1}M[h;1-z]M[f_d;z]dz
 \end{align}
 where $r$ lies in the strip of analyticity of $M[h;1-z]M[f_d;z]$ and where $M[g;s]$, is the Mellin transform
\begin{align}
M[g;s]&=\int_0^\infty x^{s-1}g(x)dx.
\end{align}
By constructing a contour that extends the line integral to an infinite rectangle around the analytic continuation of the integrand where it is no longer holomorphic, but meromorphic, along with the ability to disregard the other line integrals either generally (at $z=x\pm i\infty$) or in the $\lambda\to 0$ limit, it follows that the asymptotic terms are related to the residues of any poles of the above integrand contained in that contour \cite{norman_bleistein_asymptotic_1986}. Owing to the $(2\pi i)^{-1}$ factor the expansion is then expressed as those residues contained in the contour multiplied by $-1$. Since $h=e^{-t}$ it follows that $M[h;1-s]=\Gamma[1-s]$ such that it is non-analytic at  $s=1,2,3,\ldots$ (i.e. at $s=a_i+1$), whilst the $p$-test of $f(t)$ indicates a pole in $M[f_d;s]$ at $s=d/2$. Since we have the asymptotic expressions in Eq.~(\ref{asym}) which describes the divergent part of the integrals, we have the Laurent series for these poles which can be expressed
\begin{align}
M[f_d;s]_{s=\frac{d}{2}}&=-\frac{(2\pi)^{-\frac{d}{2}}}{(s-\frac{d}{2})}\nonumber\\
M[h;1-s]_{s=a_i+1}&=-\frac{(-1)^{i}}{i!(s-a_i-1)}.
\end{align}
If the poles in $M[f_d;s]$ and $M[h;1-s]$ do not coincide then the residues are of order $1$ and can be expressed
\begin{align}
&-\text{res}\left\{\lambda^{z-1}M[h;1-z]M[f_d;z]\right\}\nonumber\\
&=-M[f_d;s]_{s=\frac{d}{2}}\left[\lambda^{z-1}M[h;1-s]\right]_{s\to\frac{d}{2}}\nonumber\\
&\qquad-\sum_{i=0}^\infty M[h;1-s]_{s=a_i+1}\left[ \lambda^{z-1}M[f_d;z]\right]_{z\to a_i+1}\nonumber\\
&=\lambda^{\frac{d}{2}-1}(2\pi)^{-\frac{d}{2}}M\left[h;1-\frac{d}{2}\right]+\sum_{i=0}^\infty \lambda^{a_i}\frac{(-1)^i}{i!}M[f_d;a_i+1]\nonumber\\
&=\lambda^{\frac{d}{2}-1}(2\pi)^{-\frac{d}{2}}\Gamma\left[1-\frac{d}{2}\right]+\sum_{i=0}^\infty \lambda^{i}\frac{(-1)^i}{i!}M[f_d;i+1]
\label{exp1}
\end{align}
which is always the case for odd $n$. For even $n$ there are values of $a_i$ for which $M[h;1-z]M[f_d;z]$ is a double pole, namely when $z=a_i+1=d/2$. For this case the residue is written
\begin{widetext}
\begin{align}
\underset{z\to\frac{d}{2}}{\text{res}}\left\{\lambda^{z-1}M[h;1-z]M[f_d;z]\right\}&=\lambda^{\frac{d}{2}-1}\underset{z\to\frac{d}{2}}{\text{res}}\left\{\left(1+\left(z-\frac{d}{2}\right)\ln{(\lambda)}+\ldots\right)M[h;1-z]M[f_d;z]\right\}
\end{align}
which is a second order pole and a first order pole respectively (followed by higher order terms in the expansion which do not lead to poles), such that we then write
\begin{align}
&-\lambda^{\frac{d}{2}-1}\underset{z\to\frac{d}{2}}{\text{res}}\left\{\left(1+\left(z-\frac{d}{2}\right)\ln{(\lambda)}+\ldots\right)M[h;1-z]M[f_d;z]\right\}\nonumber\\
&\quad=-\lambda^{\frac{d}{2}-1}\ln{(\lambda)}\lim_{z\to\frac{d}{2}}\left[\left(z-\frac{d}{2}\right)^2M[h;1-z]M[f_d;z]\right]-\lambda^{\frac{d}{2}-1}\lim_{z\to\frac{d}{2}}\frac{d}{dz}\left[\left(z-\frac{d}{2}\right)^2M[h;1-z]M[f_d;z]\right]\nonumber\\
&\quad=-\lambda^{\frac{d}{2}-1}\ln{(\lambda)}(2\pi)^{-\frac{d}{2}} \frac{(-1)^{\frac{d}{2}-1}}{\left(\frac{d}{2}-1\right)!}-\lambda^{\frac{d}{2}-1}\lim_{z\to\frac{d}{2}}\frac{d}{dz}\left[\left(z-\frac{d}{2}\right)^2M[h;1-z]M[f_d;z]\right]\nonumber\\
&\quad=-\lambda^{\frac{d}{2}-1}\ln{(\lambda)}(2\pi)^{-\frac{d}{2}} \frac{(-1)^{\frac{d}{2}-1}}{\left(\frac{d}{2}-1\right)!}\nonumber\\
&\qquad-\lambda^{\frac{d}{2}-1}\lim_{z\to\frac{d}{2}}\left[\left(\left(z-\frac{d}{2}\right)M[h;1-z]\right)\frac{d}{dz}\left(\left(z-\frac{d}{2}\right)M[f_d;z]\right)+\left(\left(z-\frac{d}{2}\right)M[f_d;z]\right)\frac{d}{dz}\left(\left(z-\frac{d}{2}\right)M[h;1-z]\right)\right]\nonumber\\
&\quad=-\lambda^{\frac{d}{2}-1}\ln{(\lambda)}(2\pi)^{-\frac{d}{2}} \frac{(-1)^{\frac{d}{2}-1}}{\left(\frac{d}{2}-1\right)!}+\lambda^{\frac{d}{2}-1}\lim_{z\to\frac{d}{2}}\left[\frac{(-1)^{\frac{d}{2}-1}}{\left(\frac{d}{2}-1\right)!}\frac{d}{dz}\left(\left(z-\frac{d}{2}\right)M[f_d;z]\right)+(2\pi)^{-\frac{d}{2}}\frac{d}{dz}\left(\left(z-\frac{d}{2}\right)M[h;1-z]\right)\right]\nonumber\\
&\quad=-\lambda^{\frac{d}{2}-1}\ln{(\lambda)}(2\pi)^{-\frac{d}{2}} \frac{(-1)^{\frac{d}{2}-1}}{\left(\frac{d}{2}-1\right)!}+\lambda^{\frac{d}{2}-1}\frac{(-1)^{\frac{d}{2}}\left(\gamma-H_{\frac{d}{2}-1}\right)}{(2\pi)^{\frac{d}{2}}\left(\frac{d}{2}-1\right)!}+\lambda^{\frac{d}{2}-1}\lim_{z\to\frac{d}{2}}\left[\frac{(-1)^{\frac{d}{2}-1}}{\left(\frac{d}{2}-1\right)!}\frac{d}{dz}\left(\left(z-\frac{d}{2}\right)M[f_d;z]\right)\right]
\end{align}
where $\gamma\simeq 0.577$ is the Euler Gamma constant and $H_n=\sum_{i=1}^n i^{-1}$, ($H_0=0$), is the $n$-th harmonic number. Manipulations come from directly substituting the relevant terms from the Laurent series, differentiating by parts, applying $\lim a\cdot b=\lim a\cdot\lim b$ and then recognising $M[h;1-z]=\Gamma[1-z]$. These terms then replace the corresponding omitted terms corresponding to $z=a_i+1=d/2$ in Eq.~(\ref{exp1}). Importantly, when paired with the pre-factor from Eq.~(\ref{ltform}), for even $d>4$, all terms from this contribution are slower than $(p-1/2)^2$.
\par
Having established the asymptotic forms for even and odd $d$, we wish to consider the leading order term for each $d$. To so we first consider the leading term in Eq.~(\ref{exp1}) which only contributes for odd $d$. In these cases we find (assuming $p\geq 1/2$ for brevity)
\begin{align}
(\lambda+d)(2\pi)^{-d/2}\Gamma[1-d/2]\lambda^{\frac{d}{2}-1}&=\frac{d}{2\sqrt{p(1-p)}}(2\pi)^{-d/2}\Gamma[1-d/2]\left(\frac{d}{2\sqrt{p(1-p)}}-d\right)^{\frac{d}{2}-1}\nonumber\\
&=\left(\frac{d}{\pi}\right)^{\frac{d}{2}}\frac{\Gamma[1-d/2]}{2}\left(p-\frac{1}{2}\right)^{d-2}+\mathcal{O}\left(\left(p-\frac{1}{2}\right)^d\right)
\end{align}
allowing us to identify contributing terms in $u(1,p)\sim (1/2)(p-1/2)^{-1}$, $u(3,p)\sim-3\sqrt{3}\pi^{-1}(p-1/2)$, $u(5,p)\sim 50\sqrt{5}(3\pi^2)^{-1}(p-1/2)^3$, \ldots.
\par
Next, for even $d$ we must consider the contribution from the double pole at $a_i+1=d/2$. The leading order expression is that which contains the $\ln(\lambda)$ term which is finally
\begin{align}
\frac{(-1/(2\pi))^{\frac{d}{2}}}{(\frac{d}{2}-1)!}(\lambda+d)\lambda^{\frac{d}{2}-1}\ln(\lambda)&=\frac{(d/\pi)^{d/2}}{2(\frac{d}{2}-1)!}\left(p-\frac{1}{2}\right)^{d-2}\ln\left[2d\left(p-\frac{1}{2}\right)^2\right]
\end{align}
\end{widetext}
where we note a term of order $(p-1/2)^{d-2}$ also contributes.
\par
Next, we address the sum in Eq.~(\ref{exp1}) and note that
\begin{align}
M[f_d,1+i]&=\int_0^{\infty}x^i\left[e^{-x}I_0(x)\right]^ddx
\end{align}
which by the above asymptotic arguments only converges for $i<(d/2)-1$ which is to be interpreted as its valid domain, which by comparison with the double pole condition $z=a_i+1=d/2$ ensures that it contributes, where it exists, for all $d$, odd and even. Considering the leading order contributions from this sum we understand that we have, again including the relevant pre-factor,
\begin{align}
(\lambda+d)\lambda^i&=\frac{d}{2\sqrt{p(1-p)}}d^i\left(\frac{1}{2\sqrt{p(1-p)}}-1\right)^i\nonumber\\
&=\begin{cases}
      \mathcal{O}(1)+\mathcal{O}\left(\left(p-\frac{1}{2}\right)^2\right), & \text{if}\ i=0, n>2\\
      \mathcal{O}\left(\left(p-\frac{1}{2}\right)^2\right), & \text{if}\ i=1, n> 4\\
      \mathcal{O}\left(\left(p-\frac{1}{2}\right)^4\right), & \text{if}\ i=2, n> 6 \\
      \mathcal{O}\left(\left(p-\frac{1}{2}\right)^6\right), & \text{if}\ i=3, n> 8\\
      \ldots
    \end{cases}
    \label{mellin_i}
\end{align}
revealing that the sum of such terms only contribute at most quadratically in $(p-1/2)$, less a constant that appears as expected for $n\geq 3$ when $u(d,1/2)\neq \infty$. One can use this to then confirm the form of $u(d,1/2)$ and the correctness of the quadratic approximation for $d\geq 5$ from the above. The $\mathcal{O}(1)$ term in $(p-1/2)$ here, contributing only for $d\geq 3$ is given by
\begin{align}
u(d,1/2)&=dM[f_d,1]\nonumber\\
&=d\int_0^\infty\left[e^{-x}I_0(x)\right]^ddx\nonumber\\
&=\int_0^\infty e^{-x}\left[I_0\left(\frac{x}{d}\right)\right]^ddx
\end{align}
which is the Montroll extension of the Poly\'a result as expected \cite{montroll_random_1956}. In turn the quadratic coefficient in $u(d,p)$ from the above analysis and is found to be
\begin{align}
&2dM[f_d,1]-2d^2M[f_d,2]\nonumber\\
&=2\int_0^\infty e^{-x}\left[I_0\left(\frac{x}{d}\right)\right]^ddx-2\int_0^\infty xe^{-x}\left[I_0\left(\frac{x}{d}\right)\right]^ddx\nonumber\\
&=2\int_0^\infty e^{-x}\left[I_0\left(\frac{x}{d}\right)\right]^ddx-\left[2e^{-x}(-1-x)\left[I_0\left(\frac{x}{d}\right)\right]^d\right]_0^\infty\nonumber\\
&\quad-2\int_0^\infty (1+x)e^{-x}\left[I_0\left(\frac{x}{d}\right)\right]^{d-1}I_1\left(\frac{x}{d}\right)dx\nonumber\\
&=-2\int_0^\infty xe^{-x}\left[I_0\left(\frac{x}{d}\right)\right]^{d-1}I_1\left(\frac{x}{d}\right)dx
\end{align}
since $(d/dx)I_0(x)=I_1(x)$ which then matches the naive Taylor expansion result in Eq.~(\ref{sd}), confirming its suitability for $d\geq 5$ (though it does not confirm the validity of the Taylor series expansion), with the quadratic coefficient in work extracted for the optimal demon in turn being given by
\begin{align}
\frac{T\mathcal{W}(e^{-1})\left(2d^2M[f_d,2]-2dM[f_d,1]\right)}{\left(dM[f_d,1]\right)^2}.
\label{quad}
\end{align}
\par
Finally, we can use these asymptotics to consider the leading terms in $p-1/2$ in the work extracted (less $\mathcal{O}(1)$ terms). In summary, for both odd and even $d$ there is a contribution of order $(p-1/2)^{d-2}$, whilst for even $d$ there is an additional contribution of order $(p-1/2)^2\ln(\sqrt{2d}(p-1/2))$. In addition to these contributions there is an $\mathcal{O}(1)$ and $(p-1/2)^2$ contribution for $d>2$, with an additional $(p-1/2)^2$ contribution appearing for $d>5$. 
\par
Since $P_u^s(d,p)=u^{-1}(d,p)$, for $d=1$ and $d=2$ where there are no $\mathcal{O}(1)$ terms the leading order term in $P_u^s(d,p)$ goes as the inverse leading order term in $u(d,p)$, whereas for $d\geq 3$ the leading order terms go as $\mathcal{O}(1)$ followed by a term proportional to the (negative) next leading term in $u(d,p)$. To summarise we have the following asymptotic behaviour in $u(d,p)$ and $P_u^s(d,p)$, abbreviating $\Delta p=(p-1/2)$
\begin{widetext}
\begin{center}
\begin{tabular}{|c|c|c|}
\hline
$d$&$u(d,1/2+\Delta p)$&$P_u^s(d,1/2+\Delta p)$\\
\hline
$1$&$\mathcal{O}(\Delta p^{-1})$&$\mathcal{O}(\Delta p)$\\
$2$&$\mathcal{O}(\ln (2\Delta p))$&$\mathcal{O}(1/\ln (2\Delta p))$\\
$3$&$\mathcal{O}(1)+\mathcal{O}(\Delta p)$&$\mathcal{O}(1)+\mathcal{O}(\Delta p)$\\
$4$&$\mathcal{O}(1)+\mathcal{O}((\Delta p)^2\ln (2\sqrt{2}\Delta p))$&$\mathcal{O}(1)+\mathcal{O}((\Delta p)^2\ln (2\sqrt{2}\Delta p)$\\
$\geq 5$&$\mathcal{O}(1)+\mathcal{O}(\Delta p^2)$&$\mathcal{O}(1)+\mathcal{O}(\Delta p^2)$\\
\hline
\end{tabular}
\end{center}
\end{widetext}
This finally allows us to conclude that, because near $p=1/2$ the cost of biasing the exploration of the storage medium can be approximated as $8T(p-1/2)^2$ and that from Eq.~(\ref{d5work}) we have established that for the quadratic term for $d\geq 5$ is well below this value, deviations in $p$ around $p=1/2$ are \emph{always} sub-optimal for $d\geq 5$ and that, conversely, for $d\leq 4$ there is \emph{always} a deviation away from $p=1/2$ which is beneficial so long as the demon has been properly set up (such that the random access model is capable of work extraction) demonstrating the claim of a second critical dimension $d=5$.
\par
Finally, we remark on the case $d=4$. Numerically computing $u(4,p)$ indicates an optimal $p$ of $0.5$ down to machine precision, however the asymptotic analysis points towards a maximum away from $p=0.5$. The mathematical, if not numerically significant, maximum can be illustrated through explicit expansion of terms. For $d=4$ we have
\begin{align}
&u(4,p)\nonumber\\
&\sim 4M[f_4,1]+\frac{8}{\pi^2}\ln\left[8(p-1/2)^2\right](p-1/2)^2\nonumber\\
&+\left[8M[f_4,1]+\frac{8}{\pi^2}(\gamma-1)-32\underbrace{\lim_{z\to 2}\frac{d}{dz}(z-2)M[f_4,z]}_{c}\right]\nonumber\\
&\quad\times(p-1/2)^2
\end{align}
where $\gamma\simeq 0.577$ is the Euler Gamma constant. The last limit term, $c$, is not obtainable analytically and difficult to numerically estimate with much precision, but simulation points towards a bound  $c< b\simeq 0.1$. By differentiating the maximum net work, $\mathcal{W}(e^{-1})T/u(d,p) - (2p-1)T\ln(p/(1-p))$, and expanding to $\mathcal{O}((p-1/2)^2)$  we can solve for its stationary point, the optimal value of $p=p^d_{\text{opt}}$, which gives
\begin{align}
&p^4_{\text{opt}}\simeq\frac{1}{2}\nonumber\\
&+2^{-3/2}e^{\left[-\frac{1}{2}\left(\gamma+\pi^2\left[M[f_4;1]-4c+16(\mathcal{W}(e^{-1})^{-1}(M[f_4;1])^2\right]\right)\right]}.
\label{4opt}
\end{align}
Choosing $c=0.1$ (which corresponds to a maximum estimate of $p_{\text{opt}}^4$) gives $p_{\text{opt}}^4\sim 0.5+6\times 10^{-13}$ which is far below a value which can be distinguished from the $d\geq 5$ result computationally.
\section{Relation of the critical dimensions to transience of random walkers}
\label{smIV}
Here we make explicit the connection between the transience of the random walker to the critical dimensions. Initially we consider the first critical dimension which concerns work extraction at zero bias. Since the long term ability of a random walk to discover new sites is related to the transience of a random walk the first dimension directly follows the well known result that at zero bias, for $d<2$ random walks are recurrent and for $d\geq 3$ random walks are transient \cite{polya_uber_1921,b._c._hughes_random_1996}.
\par
Next we consider the second critical dimension. Here we consider whether, at zero bias, a random walker is \emph{strongly} transient, which is to be understood the mean return time to the origin is finite \cite{b._c._hughes_random_1996}. This occurs when the following limit
\begin{align}
\lim_{z\to 1^-}\frac{d}{dz}G(\mathbf{o},z)\Big|_{p=1/2}
\end{align}
converges. We note that we may write
\begin{align}
\left.G(\mathbf{o},z)\right|_{p=1/2}&=\int_0^\infty e^{-x}\left[I_0\left(\frac{xz}{d}\right)\right]^d dx
\end{align}
from Eq.~(\ref{genbessel}). The first derivative then has the exact same convergence properties as Eq.~(\ref{sd}), i.e. the random walker is strongly transient when $P_u^s(d,p)$ is analytic in the limit $p\to 1/2$ which occurs when $d\geq 5$. Since the leading order terms, when it is analytic, and the work spent driving the tape are equal (quadratic)  it then follows that any random walker that is \emph{not} strongly transient has an optimal bias $p_{\text{opt}}\neq 1/2$.  Having then demonstrating that the quadratic coefficient in the work spent driving the tape is greater than that extracting using it for $d\geq 5$ the general result follows: dimensions where unbiased walkers are strongly transient have optimal bias $p=1/2$ whereas dimensions where unbiased random walkers are not strongly transient have optimal bias $p\neq 1/2$.
\section{Generalisation to $d\in\mathbb{R}_+$ and approximate expressions for optimal $p$}
By recasting Eq.~(\ref{int1}) in the form found in Eq.~(\ref{int2_app}) we note that we have extended the $d$-domain of $P_u^s(d,p)$ to the entire positive real line, perhaps opening up the possibility of exploration of tape symbols on fractal structures \cite{b._c._hughes_random_1996}, though this is beyond scope here. However, a simple generalisation of the above yields results for the regions of $d$ for which the distinct behaviour associated with each side of the critical dimensions occurs. 
\par
\begin{figure}
  \caption{Illustration of the first critical dimension through variation of, $P_u^s(d,1/2)$, as a function of dimensionality $d$, $d\in\mathbb{R}_+$, with no work extraction possible at zero bias for $d\leq 2$, but work extraction possible for $d>2$.}
  \centering
    \includegraphics[width=0.47\textwidth]{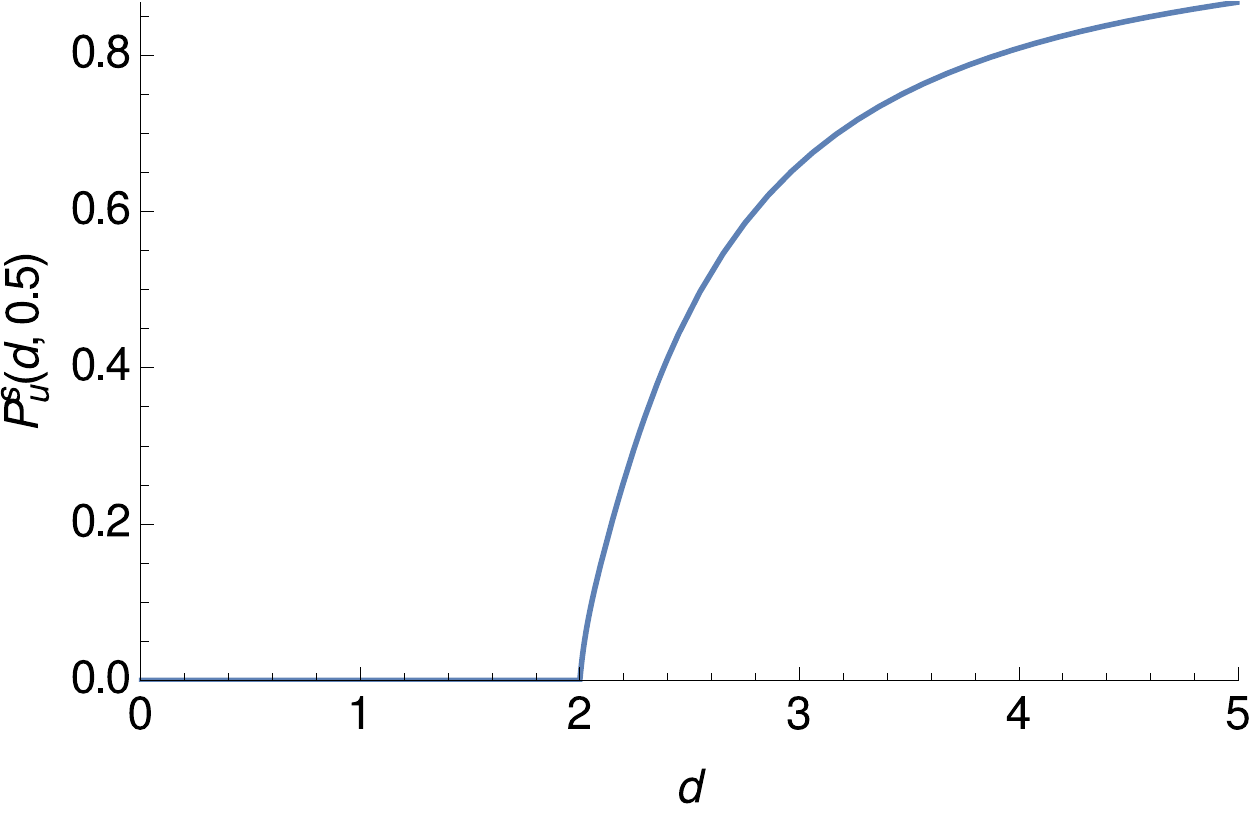}
    \label{dc1}
\end{figure}

\begin{figure}
  \caption{Optimal value of bias, $p_{\text{opt}}$, as a function of dimensionality $d$, $d\in\mathbb{R}_+$, close to the second critical dimension. To the right of the dashed line the optimal bias is always $0$ ($p=1/2$). Behaviour calculated with the approximation in Eqs.~(\ref{approx})-(\ref{approx3})}
  \centering
    \includegraphics[width=0.47\textwidth]{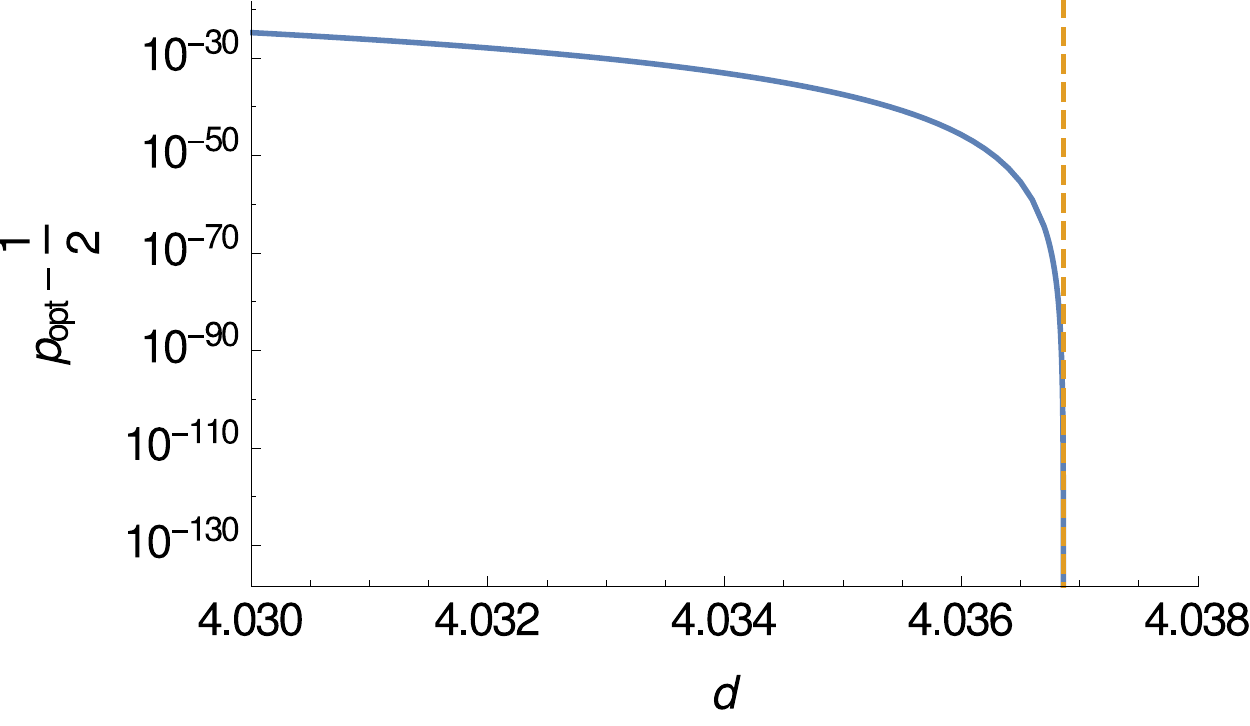}
    \label{dc2}
\end{figure}

 
In this general picture, work cannot be extracted at zero bias for $d\in[0,2]$ whilst it can for $d\in(2,\infty)$ such that we recognise the first critical value of $d\in\mathbb{R}_+$ to be $d_{\text{crit}}^1=\sup [0,2]=2$. Similarly, if the random access demon can extract work at zero bias, the interval in $d$ for which some bias is optimal is $d\in(0,4+\delta)$ where $0<\delta<1$ such that $d_{\text{crit}}^2=\sup (0,4+\delta)=4+\delta$. For an optimal demon $\delta$ takes its maximum value when the quadratic coefficient in Eq.~(\ref{quad}) equals the quadratic coefficient in the cost of driving the tape, i.e. it is the solution to
\begin{align}
&8=2\mathcal{W}(e^{-1})\left((4+\delta_{\text{max}})M[f_{4+\delta_{\text{max}}};1]\right)^{-2}(4+\delta_{\text{max}})\nonumber\\
&\quad\times\left((4+\delta_{\text{max}})M[f_{4+\delta_{\text{max}}};2]-M[f_{4+\delta_{\text{max}}};1]\right)
\end{align}
from which we estimate $\delta_{\text{max}}\sim 0.036949$ and observe that as a demon is made less optimal (but still capable of work extraction), $\delta$ approaches $0$, but never passes it. This goes some way to explaining the asymptotic behaviour at $d=4$ as it is demonstrably the infimum in $d_{\text{crit}}^2$ that divides the two qualitative behaviours, i.e. $d^2_{\text{crit}}\in(4,4+\delta_{\text{max}}]$. These two critical behaviours are illustrated in Figs.~(\ref{dc1}) and (\ref{dc2}). It is also worth noting that achieving these plots is generally not feasible numerically directly, and so approximation strategies typically need to be employed. For instance for $2<d\lesssim 2.7$ the integral that defines $P_u^s(d,1/2)$ is dominated by a fat tail which makes convergence difficult. As such one can reproduce such behaviour, for $d>2$, through the following approximation
\begin{align}
&\int_0^\infty e^{-z}\left[I_0\left(\frac{z}{d}\right)\right]^d dz\nonumber\\
&=\int_0^c e^{-z}\left[I_0\left(\frac{z}{d}\right)\right]^d dz+\int_c^\infty e^{-z}\left[I_0\left(\frac{z}{d}\right)\right]^d dz\nonumber\\
&=\lim_{c\to \infty}\int_0^c e^{-z}\left[I_0\left(\frac{z}{d}\right)\right]^d dz+\int_c^\infty e^{-z}\left[\frac{e^{\frac{z}{d}}}{\sqrt{2\pi \frac{z}{d}}}\right]^d dz\nonumber\\
&=\lim_{c\to \infty}\int_0^c e^{-z}\left[I_0\left(\frac{z}{d}\right)\right]^d dz+\frac{2^{1-d/2}c^{1-d/2} d^{d/2}\pi^{-d/2}  }{d-2}
\end{align}
since $I_0(x)\sim e^{x}/\sqrt{2\pi x}$ for large $x$. This then replicates the behaviour well given sufficiently large values of $c$. Similar strategies can be employed in the computation of Mellin transforms of $f_d$.
\par
\begin{figure}
  \caption{Optimal value of bias, $p_{\text{opt}}$, as a function of dimensionality $d$, $d\in\mathbb{R}_+$, with approximate forms.}
  \centering
    \includegraphics[width=0.47\textwidth]{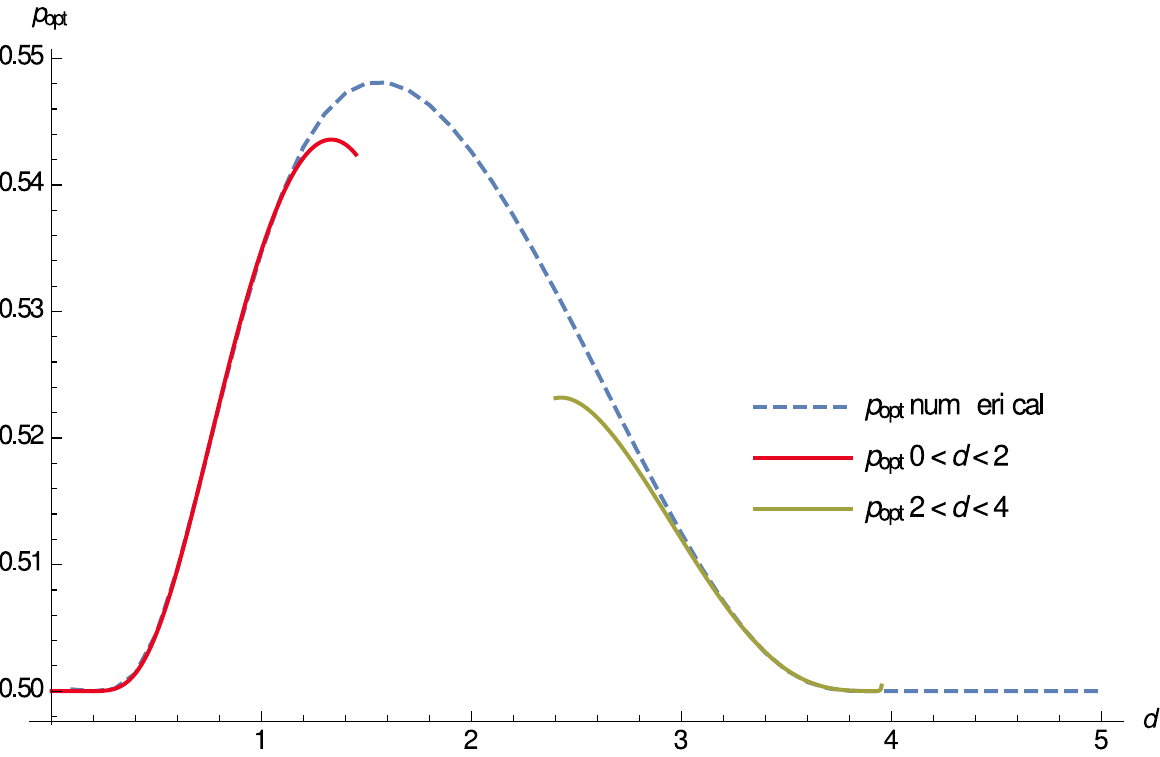}
    \label{optp}
\end{figure}

Moreover, in keeping with the generalisation to $d\in\mathbb{R}_+$, we can approximate closed form analytical solutions for $p^d_{\text{opt}}$, as opposed to the ad hoc (albeit more precise) approximation utilised in Eq.~(\ref{4opt}), in different domains of validity in order to then describe them for given (integer) $d$. For instance, by taking leading order terms in the work extracted less the work spent on driving the tape, differentiating and solving for the maxima we find the following approximations (noting that logarithmic terms at $d=2,4$ would require further individual treatment), for $0<d<2$
\begin{align}
p^{0<d<2}_{\text{opt}}&\simeq\frac{1}{2}+\frac{\mathcal{W}(e^{-1})}{8}\left[\left(\frac{\pi}{d}\right)^{\frac{d}{2}}\frac{(d-2)}{\Gamma\left[1-\frac{d}{2}\right]}\right]^{d^{-1}},
\label{approx}
\end{align}
for $d<2<4$,
\begin{align}
&p^{2<d<4}_{\text{opt}}\simeq\frac{1}{2}\nonumber\\
&+\left[\left(\frac{d}{\pi}\right)^{\frac{d}{2}}\frac{\mathcal{W}(e^{-1})\Gamma\left[2-\frac{d}{2}\right]}{16(dM[f_d;1])^2+4\mathcal{W}(e^{-1})dM[f_d;1]}\right]^{\frac{1}{4-d}},
\end{align}
and $4<d<4+\delta_{\text{max}}$
\begin{align}
&p^{4<d<4+\delta_{\text{max}}}_{\text{opt}}\simeq\frac{1}{2}\nonumber\\
&+\left[\frac{\mathcal{W}(e^{-1})(d-2)d^{\frac{d}{2}-2}\pi^{-\frac{d}{2}}\Gamma\left[1-\frac{d}{2}\right]}{M^2[f_d;1]\left(\frac{2\mathcal{W}(e^{-1})\left(2d^2M[f_d;2]-2dM[f_d;1]\right)}{\left(dM[f_d;1]\right)^2}-16\right)}\right]^{\frac{1}{4-d}},
\label{approx3}
\end{align}
with $p^{d>4+\delta_{\text{max}}}_{\text{opt}}=\frac{1}{2}$ for $d>4+\delta_{\text{max}}$. 
The first approximation performs well until $d\simeq 1.1$ where the optimal $p$ becomes too large to be described by leading order terms. Similarly the second and third approximations perform well from $d\sim 2.9$ to $d\sim 3.9$, below which it fails in the same manner as the first expression and above which second leading order terms in $(p-1/2)$ in the extracted work cannot be readily neglected since leading order terms become very similar in magnitude. This behaviour is illustrated in Fig.~(\ref{optp}).
 Fortunately they well describe the regions around $d=1$ and $d=3$ allowing us to express the following approximations
\begin{align}
p_{\text{opt}}^1&\simeq\frac{1}{2}+\frac{\mathcal{W}(e^{-1})}{8}\nonumber\\
&\simeq 0.5348\nonumber\\
p_{\text{opt}}^3&\simeq\frac{1}{2}+\frac{3\sqrt{3}\mathcal{W}(e^{-1})}{\pi\left(144(M[f_3;1])^2+12\mathcal{W}(e^{-1})M[f_3;1]\right)}\nonumber\\
&\simeq 0.5120.
\end{align}
\end{appendix}

\bibliographystyle{apsrev4-1}

\end{document}